\documentclass[%
superscriptaddress,
preprint,
nofootinbib,
amsmath,amssymb,
aps,
pra,
floatfix,
]{revtex4-2}

\usepackage{graphicx}
\usepackage{dcolumn}
\usepackage{bm}
\usepackage{floatrow}
\usepackage{comment}
\usepackage{mathrsfs}
\usepackage{gensymb}
\usepackage{xcolor}
\usepackage{hyperref}
\usepackage{cleveref}

\usepackage{newtxtext}
\usepackage{newtxmath}
\usepackage{natbib}
\usepackage{hyperref}
\usepackage{gensymb}
\usepackage{comment}
\usepackage{soul}
\usepackage{dirtytalk}
\usepackage{setspace}

\crefname{section}{§}{§§}
\Crefname{section}{§}{§§}
\hypersetup{
  colorlinks   = true, 
  urlcolor     = blue, 
  linkcolor    = blue, 
  citecolor   = blue 
}

\begin{document}
\footnote{\textbf{Notice:} This manuscript has been authored in part by UT-Battelle, LLC, under contract DE-AC05-00OR22725 with the US Department of Energy (DOE). The US government retains and the publisher, by accepting the article for publication, acknowledges that the US government retains a nonexclusive, paid-up, irrevocable, worldwide license to publish or reproduce the published form of this manuscript, or allow others to do so, for US government purposes. DOE will provide public access to these results of federally sponsored research in accordance with the DOE Public Access Plan (\href{http://energy.gov/downloads/doe-public-access-plan}{http://energy.gov/downloads/doe-public-access-plan}).}
\preprint{}

\title{Vortical interactions in turbulent thermoacoustic systems}

\author{Ankit Sahay}\email{ankitsahay02@gmail.com}
\affiliation{Department of Aerospace Engineering, Indian Institute of Technology Madras, Chennai 600036, India}
\affiliation{Centre of Excellence for studying Critical Transitions in Complex Systems, Indian Institute of Technology Madras, Chennai 600036, India}

\author{Muralikrishnan Gopalakrishnan Meena}
\affiliation{National Center for Computational Sciences, Oak Ridge National Laboratory, Oak Ridge, TN 37831, USA}

\author{R. I. Sujith}
\affiliation{Department of Aerospace Engineering, Indian Institute of Technology Madras, Chennai 600036, India}
\affiliation{Centre of Excellence for studying Critical Transitions in Complex Systems, Indian Institute of Technology Madras, Chennai 600036, India}


\date{\today}
\begin{abstract}
This study examines the dynamics of vortical interactions and their implications for mitigating thermoacoustic instability in a turbulent combustor. The regions of intense vortical interactions are identified as vortical communities in the network space of weighted directed vortical networks constructed from two-dimensional experimental velocity data. One can expect vortical interactions in the combustor to be strongest near the moment of vortex shedding, as the shed vortices gradually weaken due to dissipation while convecting downstream. However, we show that, during the state of thermoacoustic instability, there is a non-trivial consistent phase lag of approximately $52 \degree$ between the shedding of the coherent structures from the backward-facing step and the time instant when the vortical interactions attain their local maximum value. We explain this phase lag by investigating the correlation between acoustic pressure fluctuations, spatio-temporal dynamics of coherent structures, and vortical interactions in the reaction field of the combustor. We also show the aperiodic variation of vortical interactions during the states of combustion noise and aperiodic epochs of intermittency. Furthermore, the spatio-temporal evolution of pairs of vortical communities with the maximum inter-community interactions provides insight into explaining the critical regions detected in the reaction field during the states of intermittency and thermoacoustic instability, also identified in previous studies. We further show that the most efficient suppression of thermoacoustic instability via air microjet injection is achieved when steady air jets are introduced to disrupt the maximum inter-community interactions present during the state of thermoacoustic instability.
\end{abstract}
\maketitle


\section{Introduction}
\label{sec:intro}

The ubiquitous nature of turbulence around us warrants enduring efforts to understand the dynamics of turbulence. Turbulent flows contain definable flow configurations, known as coherent structures, that retain their identity over a considerable distance downstream \citep{hussain1986coherent}. Among various research avenues, investigating large-scale coherent structures has garnered substantial attention for several decades \citep{fiedler1988coherent}. As defined by \citet{hussain1983coherent}, coherent structures represent a connected, large-scale turbulent fluid mass with phase-correlated vorticity over its spatial extent. Consistent arguments have been made that coherent structures control the dominant aspects of transport \citep{zilitinkevich1970non}, mixing \citep{yu1997role}, entrainment \citep{neamtu2019lagrangian}, and noise generation in turbulent shear flows \citep{ mankbadi1984sound}.

Large-scale coherent structures play a pivotal role in various thermofluid systems, including combustion systems exhibiting thermoacoustic instability \citep{poinsot1987vortex}. Thermoacoustic instability is commonly observed in confined combustion chambers of engineering systems such as industrial burners, gas turbine combustors, and rocket engines \citep{sujith2021thermoacoustic}. The emergence of large-scale coherent structures is one of the key mechanisms that drive thermoacoustic instability in combustion systems \citep{schadow1992combustion}. In combustors prone to vortex-driven thermoacoustic instability, the roll-up of a coherent structure near the backward-facing step entrains a quantity of unburned fresh gases that burns rapidly in a very short period of time. This sudden combustion leads to an abrupt heat release that generates an acoustic wave, which produces an acoustic disturbance with a specific time lag \citep{candel2004flame}. The acoustic perturbation causes the formation of the subsequent coherent structure in the shear layer formed close to the edge of the reactant inlet \citep{ho1984perturbed}.

Most early research on the spatio-temporal dynamics in combustion systems exhibiting thermoacoustic instability centered on examining the periodic shedding of large-scale vortical structures \citep{coats1996coherent, renard2000dynamics}. \cite{yu_daily_1991} demonstrated that both the vortex and acoustic modes coexist in a backward-facing step configuration with a choked exit. Similar characteristics were also observed in the context of thermoacoustic instability in partially premixed flames within a one-sided backward-facing step combustor \citep{desai2010effect} and a bluff body combustor \citep{sivakumar2008experimental}. \cite{hong2013examining} demonstrated the emergence and growth of large-scale structures across various phases of acoustic pressure oscillations in a laboratory-scale backward-facing step combustor. Despite being one of the leading drivers of thermoacoustic instability in turbulent combustors, coherent structures remain a necessary evil in combustor design. They play a crucial role in flame holding and enhancing reactant mixing, leading to improved combustion efficiency. This underscores the ongoing need to deepen our understanding of the diverse roles played by coherent structures in turbulent combustors.

When studying the dynamics of coherent structures in the turbulent flow field of a combustor prone to thermoacoustic instability, these coherent structures rarely exist alone. Instead, various factors such as acoustic standing waves, flame stabilisation mechanisms, and combustor boundaries often influence their behaviour. One of these crucial factors is the presence of multiple coherent structures in close vicinity in a turbulent flow field. These multiple coherent structures mutually interact with each other through vortical interactions. Following the shedding of vortices from the backward-facing step, the ensuing vortical interactions assume importance as the shed vortices propagate downstream. Investigating these vortical interactions is crucial for understanding the rich spatio-temporal dynamics of turbulent thermoacoustic systems, especially from the perspective of thermoacoustic instability control.

The presence of multiple coherent structures in the flow field of turbulent combustors, along with the corresponding presence of acoustic pressure and unsteady heat release rate fields, is a clear indicator of a complex system behaviour. A complex system is defined as a system with \textit{multiple interacting components whose behaviour cannot be inferred from the behaviour of its components} \citep{smith2003complex}. The study of unsteady reactive fluid flows present in turbulent combustors often involves complex systems with high dimensionality. This high dimensionality necessitates advanced experimental and computational techniques to capture the intricate spatio-temporal dynamics and interplay of various physical processes. For such high-dimensional systems, complex networks offer a concrete mathematical framework for quantifying interactions between elements of a system \citep{barabasi2012network}. 

In recent years, the adoption of complex networks has expanded to encompass the analysis and representation of fluid flows. The framework of complex networks has been harnessed to investigate induced velocity interactions among vortical elements \citep{nair2015network}, Lagrangian motion of fluid elements \citep{ser2015flow, hadjighasem2016spectral}, oscillator-based representation of the energy fluctuations \citep{nair2018networked}, time series of fluid-flow properties \citep{zou2019complex}, triadic interactions in turbulence \citep{gurcan2017nested, gurcan2018nested}, and the effects of perturbations on time-varying vortical flows \citep{yeh2021network}. The application of complex network methodologies has expanded to investigate a wide array of turbulent flows, including two-dimensional isotropic turbulence  \citep{taira2016network}, thermoacoustic oscillations in turbulent combustors \citep{krishnan2019emergence, krishnan_2021}, wall turbulence \citep{iacobello2018visibility}, mixing in turbulent channel flow \citep{iacobello2019lagrangian}, and isotropic magnetohydrodynamic turbulence \citep{gurcan2018nested}. \citet{iacobello2021review} and \citet{taira2022network} provide a comprehensive overview of the advancements in network-inspired analysis, modelling, and control for a variety of unsteady fluid flows such as turbulent and vortical flows.

Since coherent structures are ubiquitous in fluid flows, applying clustering techniques for identifying and analyzing vortical interactions in turbulent flows has gained significant traction in the last decade. Community detection methods have been successfully used in identifying vortical structures within turbulent flow fields \citep{murayama2018characterization} and developing reduced-order models for laminar wake flows \citep{meena2018network}. In a separate study, \citet{murayama2018characterization} leveraged the spatial distribution of vortex strengths in the turbulence network and the community structure of the vorticity field to establish that the location of local minima in pressure fluctuations correspond to the proximity of the primary hub of the turbulence network to the injector rim of the combustor. Later, \cite{kawano2023complex} extended the concept of clustering of complex networks in thermofluid systems to analyze the spatio-temporal dynamics of a model single-element rocket combustor using a spatial network constructed from flux balance of acoustic energy. The authors demonstrated that the development of spatially large communities exhibiting strong connections plays a crucial role in driving thermoacoustic instability. The cumulative body of research conducted thus far has progressively deepened our comprehension of the intricate spatio-temporal dynamics of thermoacoustic systems.

Previous studies have predominantly concentrated on the shedding dynamics of coherent structures and their interactions with unsteady flame behaviour. However, there is little discussion on the dynamics of vortical interactions in turbulent thermoacoustic systems. The interaction among vortical structures has been investigated only for non-reacting flows \citep{nair2015network}. A natural initial hypothesis for reacting flows in turbulent combustors might be that vortical interactions are strongest when the vortical structures first detach from the backward-facing step. This is because the magnitude of vorticity corresponding to the detached coherent structure is at its maximum at the instant of its shedding. As these coherent structures convect downstream with the flow, their strength decreases due to the dissipation of vorticity into the surrounding flow field.

However, we show that the vortical interactions attain their maximum values at a significant distance downstream from the backward-facing step during the state of thermoacoustic instability. From the context of phase calculated for acoustic pressure oscillations during the state of thermoacoustic instability, there is a phase lag of approximately $52 \degree$ between the shedding of coherent structure from the backward-facing step and the instance when the vortical interactions attain their maximum value. To comprehensively investigate the dynamics of vortical interactions in the turbulent combustor, we employ community detection and community-based network reduction techniques on time-varying vortical networks to establish the interplay between such vortical interactions and the global acoustic dynamics exhibited by the turbulent combustor. Finally, we discuss the role of vortical interactions in the occurrence of critical regions that are suitable for open-loop control via air microjet injection. We close our discussions by demonstrating that the state of maximum suppression of thermoacoustic instability through open-loop control corresponds to the disruption of regions of vortical communities with the strongest inter-community interactions.

\section{Experimental set-up}\label{sec:expt_setup}

\begin{figure}
  \captionsetup{width=1\textwidth}
  \centerline{\includegraphics[width=1\textwidth]{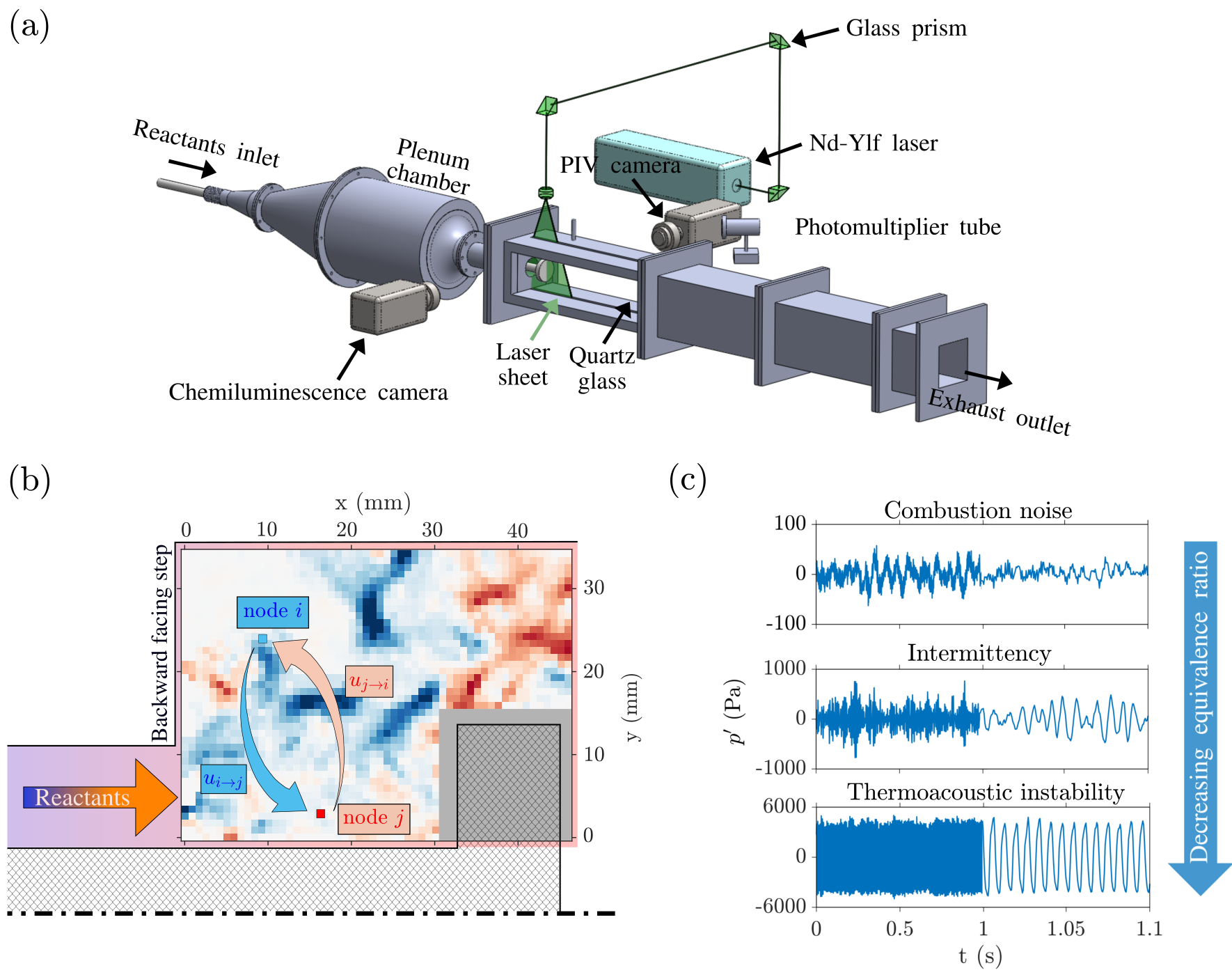}}
  \caption{(a) Schematic diagram of the experimental set-up. We simultaneously acquire the acoustic pressure measurement and high-speed two-component Particle Image Velocimetry (2C PIV) data for the current study. (b) Interactions between two vortical elements in the vorticity field were calculated from the velocity field in the turbulent combustor obtained experimentally using 2C PIV. here, $u_{i \rightarrow j}$ denotes the velocity induced by the node $i$ on node $j$, and vice-versa. The cross-hatched region indicates the bluff body and the shaft holding the bluff body. (c) Time series of acoustic pressure fluctuations during the states of combustion noise, intermittency, and thermoacoustic instability.}
\label{fig_1}
\end{figure}

Experiments were conducted in a bluff body stabilized combustor with a rectangular cross-section, possessing dimensions of 1100 $\times$ 90 $\times$ 90 mm$^3$. The combustor comprises a settling chamber, a burner section, and a rectangular duct (figure \ref{fig_1}a). In order to protect the flow entering the combustor from upstream disturbances, compressed air first enters the settling chamber. The fuel (liquefied petroleum gas - 60\% butane + 40\% propane) is injected upstream of the backward-facing step (59 mm upstream of the bluff body) through four 1 mm diameter holes bored symmetrically around the circular shaft holding the bluff body in the combustor. Thus, the fuel gets partially mixed in the air stream as it enters the combustor. We use an 11 kV ignition transformer to spark ignite the partially premixed air-fuel mixture at the backward-facing step. The flame stabilization device is a bluff body in the shape of a circular disc with a diameter of 47 mm and a thickness of 10 mm and is positioned 35 mm downstream of the backward-facing step. The combustion products are exhausted through a long duct into the atmosphere via a decoupler. The decoupler is a large chamber used to keep the combustor exit at an acoustically open boundary state ($p^\prime \approx 0$) and decouple the combustor from external perturbations.

Digital mass flow controllers (Alicat Scientific - MCR Series, 100 SLPM model for fuel flow, 4000 SLPM for the main air flow) are used to control the flow rates of air and fuel. The mass flow controllers have an uncertainty of $\pm$ (0.8 $\%$ of reading $+$ 0.2 $\%$ of full scale). In the experiments, the fuel flow rate $\dot{m}_f$ is maintained at a constant value ($\dot{m}_f = 30 \pm 0.44$ standard litre per minute (SLPM)), and the air flow rate $\dot{m}_a$ is varied ($480 \pm 7.84 \leq \dot{m}_a \leq 780 \pm 10.24$ SLPM) to change the global equivalence ratio ($\phi = (\dot{m}_f/\dot{m}_a)_{\text{actual}}/(\dot{m}_f/\dot{m}_a)_{\text{stoichiometric}}$), of the reactant mixture. The global equivalence ratio $\phi$ is the control parameter in the current study and is varied from 0.97 to 0.57. The maximum uncertainty in the calculation of $\phi$ is $\pm 0.02$. The Reynolds number (calculation based on the flow conditions of the air at the inlet of the combustor) varies in the range $19300 \leq Re \leq 31300$, with a maximum uncertainty of $\pm 400$ ($1.28$ $\%$). We measure the acoustic pressure fluctuations $p^{\prime} (t)$ using a PCB103B02 piezoelectric transducer (sensitivity: 217.5 mV/kPa and uncertainty: $\pm$0.15 Pa) mounted on the combustor wall 20 mm downstream of the backward-facing step. 

We perform high-speed two-component particle image velocimetry (2C PIV) to acquire the two components of velocity ($v_x$: streamwise parallel to the flow direction and $v_y$: perpendicular (cross-stream) to the flow direction). The reactive flow field is seeded with $\approx$ 1 mm sized TiO$_2$ particles (Kronos make product - 1071). The Stokes number ($Stk = (t_0 u_0)/l_0$) of the seeded flow, where $t_0$  is the characteristic response time of the particle, $u_0$ is the gas velocity, and $l_0$  is the characteristic dimension of the particle, is $7$ $\times$ $10^{-4}$. For such low values of $Stk$ ($Stk << 1$), the seeded particles faithfully follow the streamlines \citep{raffel1998particle}. We use a single cavity double pulsed Nd:YLF laser (with 527 nm and a pulse width of 250 ns) to illuminate the flow through a narrow quartz window (400 $\times$ 20 $\times$ 10 mm$^3$) on the top wall of the combustion chamber.

To capture the Mie scattering light from the seeding particles, we use a high-speed CMOS camera (Photron FASTCAM SA4) synchronised with the laser pulses. A short band pass optical filter (centred at 527 nm with 12 nm Full Width at Half Maximum) is mounted in front of the lens to capture the Mie scattered light. We record 3797 Mie Scattering image pairs for approximately 2 seconds. The physical region of interest used for the analysis of 2C PIV is located just downstream of the backward-facing step and slightly above the bluff body shaft, as shown in figure \ref{fig_1}(b). The Mie scattering images are processed using PIVview2C software (PIVTEC GmbH). We use the cross-correlation algorithm to calculate the velocity fields \citep{raffel1998particle}.

In the bluff body stabilised turbulent combustor considered in the present study, the dynamics exhibited by the combustor transitions from the state of combustion noise to the state of thermoacoustic instability via the state of intermittency as the global equivalence ratio of the reactant mixture is decreased from $\phi = 1$ in a quasi-static manner. Figure \ref{fig_1}(c) shows the time trace of acoustic pressure fluctuations during the three dynamical states of combustor operation, where the equivalence ratio varies from near stoichiometry to conditions corresponding to the state of thermoacoustic instability. When the equivalence ratio is close to stoichiometry, the temporal dynamics of the combustor is characterised by low-amplitude, aperiodic pressure fluctuations. This condition of combustor operation is referred to as combustion noise. The turbulent combustor exhibits the state of intermittency as the equivalence ratio is decreased. During the state of intermittency, the time series of acoustic pressure consists of epochs of high-amplitude periodic oscillations appearing in a seemingly random manner between epochs of low-amplitude aperiodic fluctuations \citep{nair2014intermittency}. The onset of thermoacoustic instability, characterised by periodic acoustic pressure oscillations of high-amplitude, occurs as the equivalence ratio is decreased further. 

Once the raw velocity data acquired during the three different dynamical states mentioned above are processed, the final velocity data corresponding to the region of interest at each time instant have the following dimensions during different states of the combustor operation:
\begin{enumerate}
      \item Combustion noise - 37.35 mm $\times$ 49.8 mm, corresponding to 45 $\times$ 60 pixels.
      \item Intermittency - 34.23 mm $\times$ 46.45 mm, corresponding to 42 $\times$ 57 pixels.
      \item Thermoacoustic instability - 34.86 mm $\times$ 46.48 mm, corresponding to 42 $\times$ 56 pixels.
      
\end{enumerate}
The different dimensions of the regions of investigation (ROIs) are attributed to experimental constraints. The primary differences between the different regions occur on the top and right sides of the flow field. The differences in ROIs during the three dynamical states do not significantly affect the results presented in the present work. The crucial vortical structures, which are central to understanding the vortical interactions during the different dynamical states exhibited by the combustor, are consistently captured within the ROIs. An in-depth elucidation of the preprocessing and post-processing algorithms employed for the analysis of 2C PIV data, as well as a comprehensive discussion concerning velocity uncertainties, can be found in \citet{george2018pattern}. Additional details are available in the supplementary material provided by \citet{krishnan2019mitigation}.

Despite the inherently three-dimensional nature of the combustor flow field in this study, we expect a reasonably axisymmetric flow during the state of thermoacoustic instability due to the longitudinal instability mode and the circular boundary of the backward-facing step. Additionally, the periodic shedding of coherent structures from the backward-facing step, a key flow dynamic relevant to this study, is accurately captured by the 2C PIV measurements. Even during states of combustion noise and intermittency, we expect the velocity data obtained from the 2C PIV measurements to qualitatively represent the velocity values in the three-dimensional flow field. The turbulent combustor is indeed a high-dimensional system with millions of degrees of freedom. The low-dimensional vorticity field calculated via 2C PIV measurements constitutes one of the many system variables that can be utilised to infer the spatio-temporal dynamics of the high-dimensional turbulent system.

\section {Complex network description of vortical interactions and community detection}\label{sec:network_description}

\citet{taira2016network} introduced a weighted complex network analysis to characterise the vorticity interactions in a two-dimensional decaying isotropic turbulent flow field obtained from direct numerical simulations. The nodes of the network are the cells in the computational grid representing the vortical elements in the flow field, and the weight of the link is the velocity induced by the vortical elements (figure \ref{fig_1}b). The induced velocities are calculated using the Biot–Savart law \citep{tietjens1957applied, malmendier2020application}: 
\begin{equation}
    u_{i \rightarrow j}=\frac{\left|\omega_i\right| \Delta S}{2 \pi\left|\boldsymbol{r}_j-\boldsymbol{r}_i\right|}, \quad i \neq j,
\end{equation}
where $u_{i \rightarrow j}$ denotes the magnitude of the velocity induced by vortical node $i$ on a vortical node $j$, $\omega_i$ denotes the vorticity value of the vortical element assumed to be present at node $i$, $\Delta S$ denotes the area of a single computational cell which is assumed to have a vortical element, and $|\boldsymbol{r}_j - \boldsymbol{r}_i|$ indicates the distance between nodes $j$ and $i$. The weighted adjacency matrix $A \in \mathbb{R}^{n \times n}$ for this vortical flow network is defined as
\begin{equation}
  A_{ij} = \left\{
    \begin{array}{ll}
      w_{ij}, & i \neq j \\[2pt]
      0,         & i = j,
    \end{array} \right.
    \label{eqn:3.3}
\end{equation}
where $w_{ij}$ is the edge weight given by
\begin{equation}
w_{i j}=\alpha u_{i \rightarrow j}+(1-\alpha) u_{j \rightarrow i},
\label{eqn:3.4}
\end{equation}
with $\alpha \in [0,1]$ being a parameter to capture the induced velocity direction to construct the vortical network. A value $\alpha = 0$ or $1$ gives a directed adjacency matrix, whereas $\alpha = 1/2$ yields a symmetric adjacency matrix corresponding to an undirected network. The statistics of the final results presented in this study show minimal variation with changes in $\alpha$, as detailed in Appendix \ref{AppendixD}. In this study, we use $\alpha = 0$ to preserve the notion of directionality of induced velocity. Adding directions to the edges helps to differentiate between the influential and influenced vortical nodes. Thus, in the present study, we define the weighted directed vortical network as
\begin{equation}
    A_{ij} = {u}_{j \rightarrow i}.
    \label{eqn:3.5}
\end{equation}

The above formulation in (\ref{eqn:3.5}) gives an asymmetric adjacency matrix with non-binary entries, representing a weighted directed network. The velocity uncertainty of 1.25\% to 2.5\% due to 2C PIV data analysis translates to an uncertainty of 0.0139 m/s to 0.0167 m/s in the edge weights of the adjacency matrix (considering a velocity of 5 m/s of the fluid particles during the state of thermoacoustic instability). The adjacency matrix contains zero values on its diagonal due to the absence of velocity influence between vortical nodes onto themselves. As every vortical node is interconnected with all others, the vortical network is a complete graph.

It is not easy to characterize the interactions between the vortical components since the spatio-temporal dynamics exhibited by the turbulent combustor is high-dimensional in nature. Analysis, simulation, and modelling of the dynamics on such large networks demand massive amounts of processing power and necessitate decreasing the computational complexity of the problem while preserving the fundamental physics. To reduce the high dimensionality of the vortical interactions, we detect communities on the time-varying vortical networks constructed for the three dynamical states. Communities are modular clusters of nodes that are densely connected. Communities are derived from decomposing the network into sub-components that are internally strongly connected but externally only weakly connected \citep{newman2004finding}. 

Modularity maximization is a technique for community detection that is frequently used to identify communities \citep{newman2004finding}. This technique seeks to optimize the modularity quality function, $M$, by dividing a network's nodes into $K$ nonoverlapping communities, $C_1, C_2, ..., C_K$. Modularity, $M$, measures the proportion of edges in a network that link nodes within the same community in excess of the expected proportion of such edges if the network had the same community structure but random connections between the nodes. To receive a higher $M$ score, the communities that a partition defines should be internally more dense than what can be predicted by chance. The community structure of a network is, therefore, assumed to be best represented by the division that gets the highest value of $M$. Modularity $M$ is given by 
\begin{equation}
M=\frac{1}{2 n_{e}} \sum_{i j}\left[A_{i j}-\gamma_{M} \frac{s_{i}^{\text{in}} s_{j}^{\text {out}}}{2 n_{e}}\right] \delta\left(c_{i}, c_{j}\right),
\label{eqn:3.6}
\end{equation}
where
\refstepcounter{equation}
$$
s_{i}^{\text {out}}=\sum_{j} A_{j i} \quad \text {and} \quad s_{i}^{\text{in}}=\sum_{j} A_{i j}.
\eqno{(\theequation{\mathit{a},\mathit{b}})}\label{eqn:3.7}
$$
Here, $s_{i}^{\text{out}}$ and $s_{i}^{\text{in}}$ denote the out- and in-strengths of a node, respectively. In (\ref{eqn:3.6}), $\gamma_{M}$ is the resolution parameter to detect the presence of large or small communities in the network \citep{reichardt2006statistical, fortunato2007resolution}, $n_{e}$ is the total number of edges in the network, $\delta\left(c_{i}, c_{j}\right)$ is the Kronecker delta, $c_i \in \hat{C_k}$ is the label of the community to which node $i$ is assigned, and $\hat{C_k}$ is the set of $k$ community in the network. Here, $k = 1,2,...,K,$ with $K$ being the total number of communities; $K$ is not a user-defined variable, and is determined by the community detection algorithm used. Various algorithms are available to detect the communities in a network \citep{fortunato2010community}.

In the present study, we use the modularity maximisation method introduced by \citet{blondel2008fast} to detect communities. We refer to these detected communities on vortical networks as \textit{vortical communities} \citep{meena2018network}. The fluid elements with high vorticity play an important role in the dynamics exhibited by the turbulent combustor. Thus, we neglect the effects of fluid elements with low vorticity. We consider only those spatial locations for nodes in vortical communities that exhibit vorticity magnitudes greater than $|\omega|$ $=$ $456$ $s^{-1}$, $380$ $s^{-1}$, and $843$ $s^{-1}$ during the states of combustion noise, intermittency, and thermoacoustic instability, respectively. Further details on calculations for these threshold values are provided in Appendix \ref{AppendixA}.

\begin{figure}
  \captionsetup{width=1\textwidth}
  \centerline{\includegraphics[width=1\textwidth]{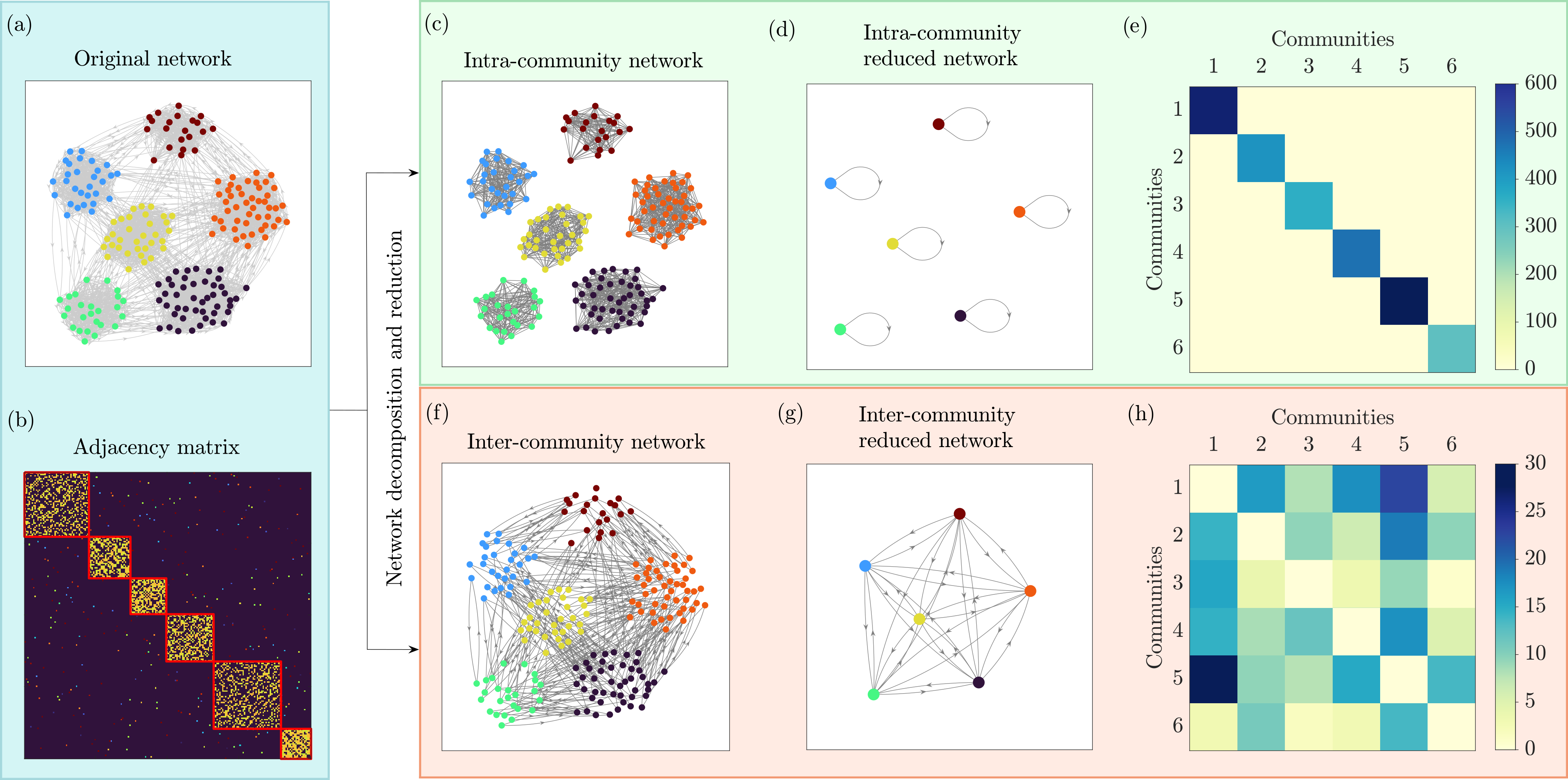}}
  \caption{An overview of community-based reduction of a representative directed complex network. (a) A random modular graph with six communities is constructed according to the Fruchterman-Reingold algorithm \citep{fruchterman1991graph}, and the corresponding adjacency matrix is shown in (b). The modular nature of the network manifests as a block structure in the adjacency matrix. The original network can be decomposed into two sub-networks: (c) an inter-community network comprising edges present only across different communities, and (f) an intra-community network comprising edges present only within the individual communities. (d,g) Each community can be considered as a single node. All the edges between a pair of communities and within each community can be summed up to obtain (d,e) a reduced inter-community network and (f,g) a reduced intra-community network. The adjacency matrix of the reduced inter-community network is shown in (h).}
\label{fig_2}
\end{figure}

We use a sample network as shown in figure \ref{fig_2} to demonstrate the dimensionality reduction method we use in the current study. In figure \ref{fig_2}, we consider a sample random modular graph constructed using methods discussed in \citet{lancichinetti2008benchmark}. Each node $i$ has a scalar weight, $\gamma_i$, associated with it. This scalar weight is analogous to the vorticity value of each fluid element in a flow system. In figure \ref{fig_2}(b), we cluster the nodes into network communities using the Louvain algorithm \citep{blondel2008fast}. The details of implementing the Louvain algorithm are given in Appendix \ref{AppendixB}. We decompose the original network into inter-community and intra-community networks in figures \ref{fig_2}(c) and  \ref{fig_2}(f), respectively. The inter-community network is further condensed to weighted-centroid-based network \citep{meena2018network} in figure \ref{fig_2}(d). The weighted centroids of each community are calculated using  
\begin{equation}
\boldsymbol{\xi}_{k} = \frac{\sum_{i \in C_{k}} \gamma_{i} \boldsymbol{r}_{i}}{\sum_{i \in C_{k}} \gamma_{i}}, \quad k=1,2, \ldots, m
\label{eqn:3.8}
\end{equation}
and the sum of all the scalar weights belonging to nodes of each community 
\begin{equation}
\Gamma_{k}=\sum_{i \in C_{k}} \gamma_{i}
\label{eqn:3.9}
\end{equation}
is concentrated at the community centroid, reducing the community to a single node at the respective community centroid. Here, $\boldsymbol{r}_{i}$ denotes the position vector of each node $i$ and $\Gamma_{k}$ is the node attribute of each node in the community-based reduced network, where $k$ refers to the $k^{\text{th}}$ community present in the original network. By employing this framework for the vortical network, we use the weight and position vector of each community centroid to reconfigure the vortical network and reduce the dimensions of the system. It is important to note that the sample network depicted in figure \ref{fig_2} is a directed network, as evident from the asymmetry of its reduced adjacency matrix shown in figure \ref{fig_2}(e). This figure is included solely to illustrate the concepts of inter-community and intra-community links and does not represent any of the directed networks analyzed in the present manuscript.

\section{Results and discussion}\label{sec:results}
We initially examine the phase-averaged vortical interactions during the state of thermoacoustic instability. The periodic nature of a self-organised flow field during this dynamical state facilitates a clear interpretation of results. The phase-averaged results obtained from the state of thermoacoustic instability also help us proceed further with the investigation of time-varying vortical networks for all the dynamical states exhibited by the combustor. 

\subsection{Phase-averaged analysis of strongest influential communities}\label{sec:phase_averaged}
\begin{figure}
  \captionsetup{width=1\textwidth}
  \centerline{\includegraphics[width=0.875\textwidth]{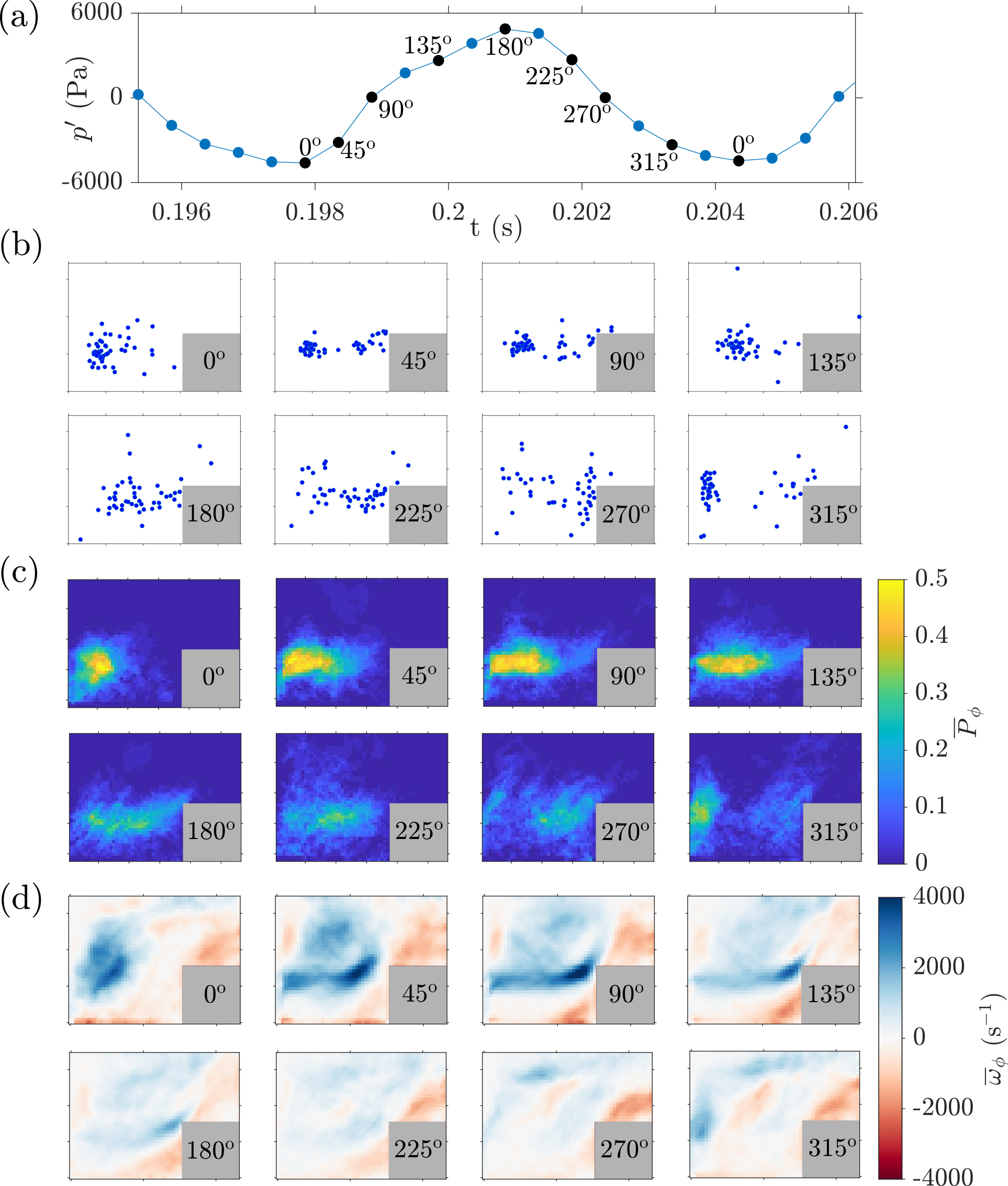}}
  \caption{(a) Time instances denoting the phases in one acoustic cycle during the state of thermoacoustic instability. (b) Shows the locations of weighted centroids of communities with maximum outward participation coefficient at each phase. (c) Phase-averaged fields of participation coefficients $P$ corresponding to the nodes in communities having the maximum $P$ values at the indicated phases of the acoustic cycle during the state of thermoacoustic instability. The participation coefficient quantifies the influence of a single node on communities other than its own. We also plot the phase-averaged distribution of vorticity in (d). The mismatch between the locations of high phase-averaged participation coefficient ($\overline{P}_{\phi}$) and high phase-averaged vorticity ($\overline{\omega}_{\phi}$) for phases $45 \degree$ to $135 \degree$ points to the inherent difference between the roles of vortical interactions and vorticity. Here, $\overline{P}_{\phi}$ reflects the combined influence of the strength and position of the vortical elements in the network space. However, $\overline{\omega}_{\phi}$ is an attribute of the individual vortical elements and does not depend on the other vortical elements.}
\label{fig_3}
\end{figure}

A significant implication of the shedding of coherent structures in unsteady fluid flows is the presence of a periodic fluctuating dynamics. The resulting periodic forces can inflict destructive, unsteady loading on the structure of the physical systems and may lead to catastrophic incidents. Since coherent structures are related to the regions of high vorticity in the hydrodynamic field, an essential step in understanding the role of interaction among regions of high vorticity toward such periodic dynamics is to use complex networks to analyze the \textit{outward} influence of vortical communities on other fluid elements. To this end, we use phase-averaged analysis during the state of thermoacoustic instability to understand the variation in the location and strength of the communities with maximum \textit{outward} influence, as shown in figure \ref{fig_3}.

We use the participation coefficient $P$ to quantify the outward influence of nodes present in vortical communities \citep{guimera2005functional}. The participation coefficient is a relative measure of the inter-community strength of a node and provides an indication of how well-distributed the edges of a node are amongst all communities other than its own. The participation coefficient $P_i$ of node $i$ is given by:
\begin{equation}
    P_{i} = 1 - \sum_k \left( \frac{s_i^k}{s_{i}} \right)^2, 
\end{equation}
\noindent where, $s_i$ is the total strength of node $i$, and $s_{i}^{k}$ is the strength of node $i$ to influence all nodes in community $k$; $s_{i}^{k}$ and $P_i$ can be evaluated for both in- and out-edges. Since we are interested in understanding the influence of a vortical community on other fluid elements, we evaluate the out-edge-based measures in the current study to understand the outward influencing properties of the fluid elements.

We plot the locations of weighted centroids of communities with maximum outward participation coefficient at each phase in figure \ref{fig_3}(b). Figure \ref{fig_3}(c) shows the phase-averaged distribution of $P_i$ of all the nodes belonging to the community with the maximum outward inter-community interaction. We use phase-averaged $P_i$ instead of calculating $P_i$ from phase-averaged vorticity fields to ensure that community affiliations that are critical for meaningful $P_i$ values are preserved and utilized to calculate $P_i$ before phase averaging is done. Using phase-averaged vorticity fields to calculate $P_i$ would smooth out essential inter-community interactions and lead to loss of information on community affiliations. Around 50 realisations for each phase of the cycle were averaged, resulting in an ensemble-averaged quantity of interest at each of the 8 phases ($\phi = 0\degree$ to $315\degree$ in intervals of $45\degree$). The phase angle $\phi=0\degree$ corresponds to the local minima of acoustic pressure fluctuations, $p^\prime$, and is shown in figure \ref{fig_3}(a). We also plot the phase-averaged distribution of vorticity $\overline{\omega}_{\phi}$ in figure \ref{fig_3}(d) to interpret the phase-averaged distribution of the participation coefficient $\overline{P}_{\phi}$ from the perspective of hydrodynamics present in the flow field during the state of thermoacoustic instability. 

We observe a region of high phase-averaged participation coefficient $\overline{P}_{\phi}$ near the backward-facing step at $\phi = 0\degree$. This region of high $\overline{P}_{\phi}$ is due to the coherent structure shedding from the backward-facing step at the same phase angle. However, this coherent structure continues to convect downstream whereas the region of high $\overline{P}_{\phi}$ continues to remain near the backward-facing step at the subsequent phase angles of $45 \degree$ to $180 \degree$. This observation is perplexing, as one would intuitively anticipate that regions characterized by high vorticity would coincide with the regions exhibiting high participation coefficients. This hypothesis arises from the understanding that regions exhibiting larger vorticity values will have larger vortical influence over other regions in the flow field.

We attribute this difference to the interpretation of participation coefficient $P_i$ from the network and vorticity $\omega$ from the points of view of the hydrodynamics. The strength of the participation coefficient $P_i$ depends on multiple factors: (i) the vorticity of the node, (ii) the size of the community to which the node belongs, and (iii) the spatial distance between the node and other communities. However, $\omega$ is an attribute of the node $i$ and does not depend on the other vortical elements. Thus, although the regions of high vorticity continue to move downstream as the phase angle increases from $0 \degree$ to $135 \degree$, the regions of highest $\overline{P}_{\phi}$ values do not spatially match with the regions of highest $\overline{\omega}_{\phi}$ values. This spatial mismatch occurs because the nodes exhibiting the strongest $\overline{P}_{\phi}$ values remain in the middle of the spatial domain (seen in figures \ref{fig_3}b and \ref{fig_3}c), as this middle region minimizes the distance to other communities, thereby enhancing the outward inter-community influence of the communities present in this region. This spatial mismatch between regions of highest $\overline{\omega}_{\phi}$ and highest $\overline{P}_{\phi}$ between the phases $45 \degree$ to $135 \degree$ indicates that, in addition to tracking the downstream convection of the coherent structure shed from the backward-facing step, spatial proximity of other regions of high vorticity must also be taken into account to understand the spatio-temporal behaviour of vortical interactions present during the state of thermoacoustic instability.

At further phase angles of $180 \degree$ to $270 \degree$, the values of participation coefficients decrease significantly, corresponding to the decrease in the vorticity values present in the flow domain. Finally, at the phase angle value of $315 \degree$, we observe the emergence of a region of high $\overline{P}_{\phi}$ values near the backward-facing step, along with an increase in the values of vorticity $\overline{\omega}_{\phi}$ in the same region. This occurs just before the coherent structures are again shed at the acoustic pressure minima ($\phi = 0 \degree$) of the next periodic cycle. The spatio-temporal dynamics of $\overline{P}_{\phi}$ and $\omega$ is subsequently repeated in all acoustic cycles during the state of thermoacoustic instability.

Values of $\overline{P}_{\phi}$ can be interpreted as an indicator of the combined influence of strength and position of the vortical elements present in the spatial field under consideration. It is possible that a fluid element with a significantly high vorticity magnitude located at the periphery of the spatial domain can exert comparatively less influence on other fluid elements than a fluid element with a lower vorticity magnitude but located towards the centre of the spatial domain. Although this information is embedded in the formulation of Biot-Savart law, and thus, in the edge weights of the adjacency matrix, $\overline{P}_{\phi}$ brings out the combined influence of the strength and position of the vortical elements attributed to various communities more clearly. As we move ahead, we examine the spatio-temporal dynamics of vortical communities to understand the reason behind the discrepancy between regions of vorticity and regions of maximum participation coefficient.

\subsection{Spatio-temporal evolution of maximum-interacting pairs of vortical communities}

In this section, we analyze the connection between the acoustic pressure oscillations, the network measures representing the spatio-temporal evolution of the maximum-interacting pair of vortical communities, and the corresponding vorticity distribution present in the reactive flow field. It is crucial to study the vorticity distribution as the vortex shedding within a turbulent combustor is closely related to the acoustic pressure fluctuations observed within the combustion chamber.

The spatio-temporal dynamics of vortical interactions is captured using tools from complex network theory discussed in Section \ref{sec:network_description}. We quantify the spatio-temporal evolution of the maximum-interacting pair of vortical communities using the average and maximum of all inter-community interactions present at any given instant of time, denoted by $A_{\text{r,mean}}$ and $A_\text{r,max}$, respectively. The subscript \textit{r} indicates that the quantities $A_{\text{r,mean}}$ and $A_\text{r,max}$ have been calculated from the reduced adjacency matrix. We compare the temporal evolution of $p^\prime$ fluctuations with the average and maximum of all the inter-community interactions during all the dynamical states exhibited by the turbulent combustor. Since the acoustic pressure remains approximately constant throughout the reaction field of the combustor \citep{george2018pattern}, tracking the temporal evolution of the average of all inter-community interactions allows us to correlate the dynamics of the entire reaction field with the acoustic pressure fluctuations. On the other hand, tracking the maximum of all the inter-community interactions will help us understand the spatio-temporal evolution of the most influential pair of vortical communities. Subsequently, this spatio-temporal evolution of vortical communities is correlated with different phases of an acoustic pressure cycle.

\subsubsection{Combustion noise}

\begin{figure}
  \captionsetup{width=1\textwidth}
  \centerline{\includegraphics[width=1\textwidth]{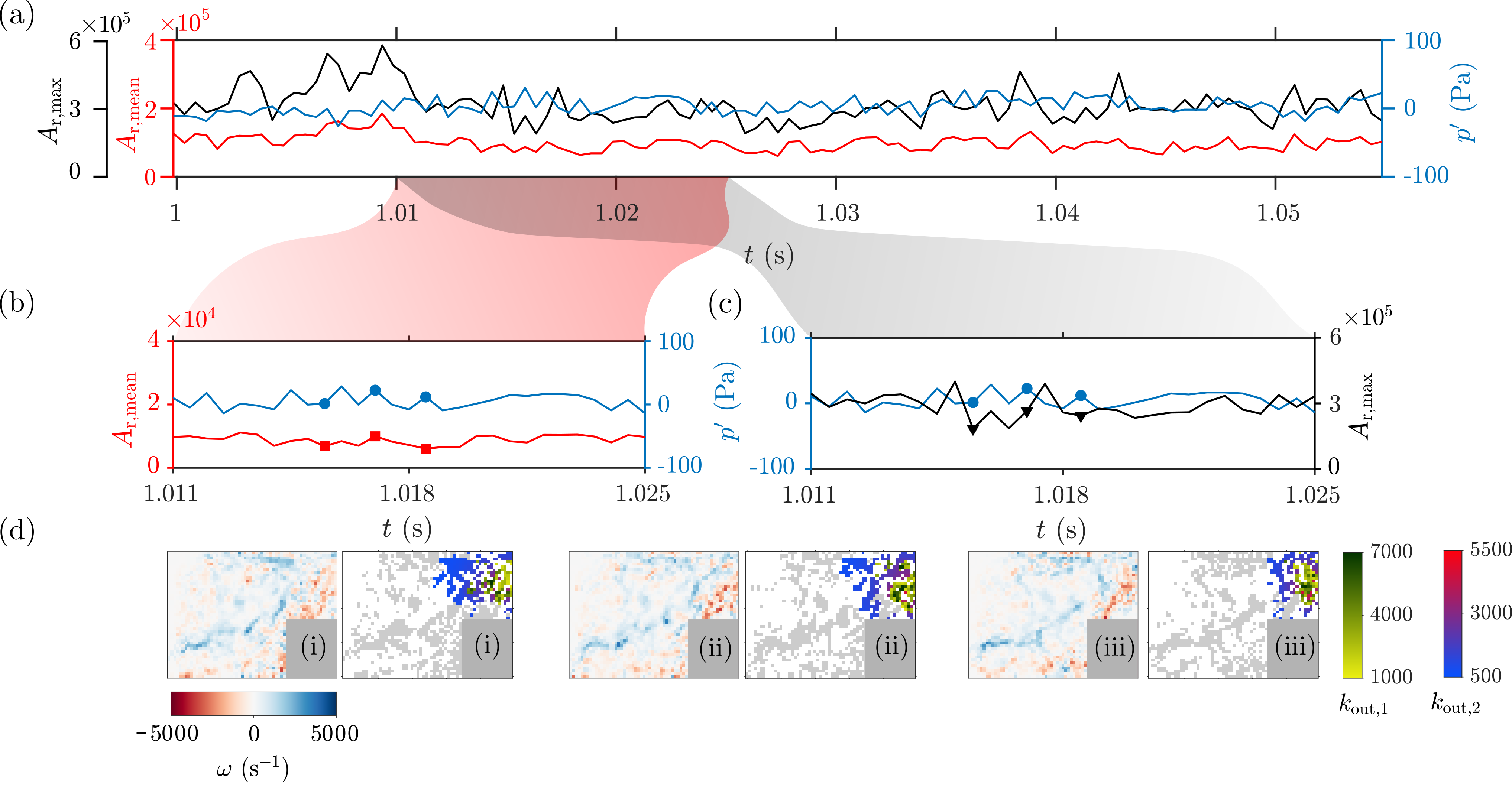}}
  \caption{(a) Time series showing the variation of $p^\prime$ fluctuations and network measures (mean $A_{\text{r,mean}}$, and maximum $A_{\text{r,max}}$ of all inter-community interactions) during the state of combustion noise. The acoustic pressure fluctuations are marked in blue. The zoomed-in versions are shown in (b-c). The two shaded regions indicate the portion in (a) used for the time series shown in (b-c). The vorticity field and the corresponding spatial distribution of the inter-community in- and out-strengths between the pair of communities with the maximum inter-community interaction at each of the representative time instants marked in (b-c) are shown in (d). The grey regions represent other vortical communities that have lower inter-community interactions. The colorbars represent the out-strengths of the pair of communities with maximum inter-community interactions, denoted by $k_{\text{out,1}}$ and $k_{\text{out,2}}$, respectively. The direction of the maximum inter-community interaction is from the community marked with $k_{\text{out,1}}$ values to the community marked with $k_{\text{out,2}}$ values. The aperiodic nature of the network measures reflects the aperiodic nature of hydrodynamic interactions present in the reaction field. Due to the shedding of significant vortices from the upstream tip of the bluff body, the pair of communities with the maximum inter-community interaction are always present above the bluff body.}
\label{fig_4}
\end{figure}

In figure \ref{fig_4}, we compare the temporal evolution of $p^\prime$ fluctuations with the average and maximum of all the inter-community interactions during the state of combustion noise. The time series of acoustic pressure oscillations and the network measures are shown in figure \ref{fig_4}(a). The zoomed-in versions are shown in figures \ref{fig_4}(b-c). We observe that the time series of network measures is aperiodic during the state of combustion noise. The aperiodic nature of the inter-community interactions results from the incoherent spatio-temporal distribution of the vortical communities present during the state of combustion noise.

In figure \ref{fig_4}(d), we show the spatio-temporal evolution of the vorticity distribution and vortical communities with the maximum inter-community interactions across three representative instants during the state of combustion noise. At each time instant, the pair of vortical communities consists of two different communities with two inter-community edges directed from, say, community 1 to community 2, and vice versa. Community 1 exhibits outward inter-community strength distribution towards community 2 with $k_{\text{out,1}}$ values, whereas community 2 exhibits outward inter-community strength distribution towards community 1 with $k_{\text{out,2}}$ values. Thus, in figure \ref{fig_4}(d), the direction of the maximum inter-community interaction is from the community marked with $k_{\text{out,1}}$ values to the community marked with $k_{\text{out,2}}$ values. 

The ranges of $k_{\text{out,1}}$ and $k_{\text{out,2}}$ are indicated in the colorbars. The spatial distribution of the pair of communities in figure \ref{fig_4}(d) indicates that most of the vortical communities with maximum inter-community interactions are located above the bluff body. This is due to a large concentration of coherent structures shed from the upstream tip of the bluff body during the state of combustion noise. Furthermore, the vortical communities shown are quite fragmented in space due to small regions of high vorticity present in the spatial domain. This contrasts the large spatially coherent vortical communities present during the aperiodic epochs of intermittency, as discussed in the next subsection. The spatial fragmentation of communities present in different dynamical states is discussed and compared in Appendix \ref{AppendixC}.

\subsubsection{Aperiodic epoch of intermittency}

\begin{figure}
  \captionsetup{width=1\textwidth}
  \centerline{\includegraphics[width=1\textwidth]{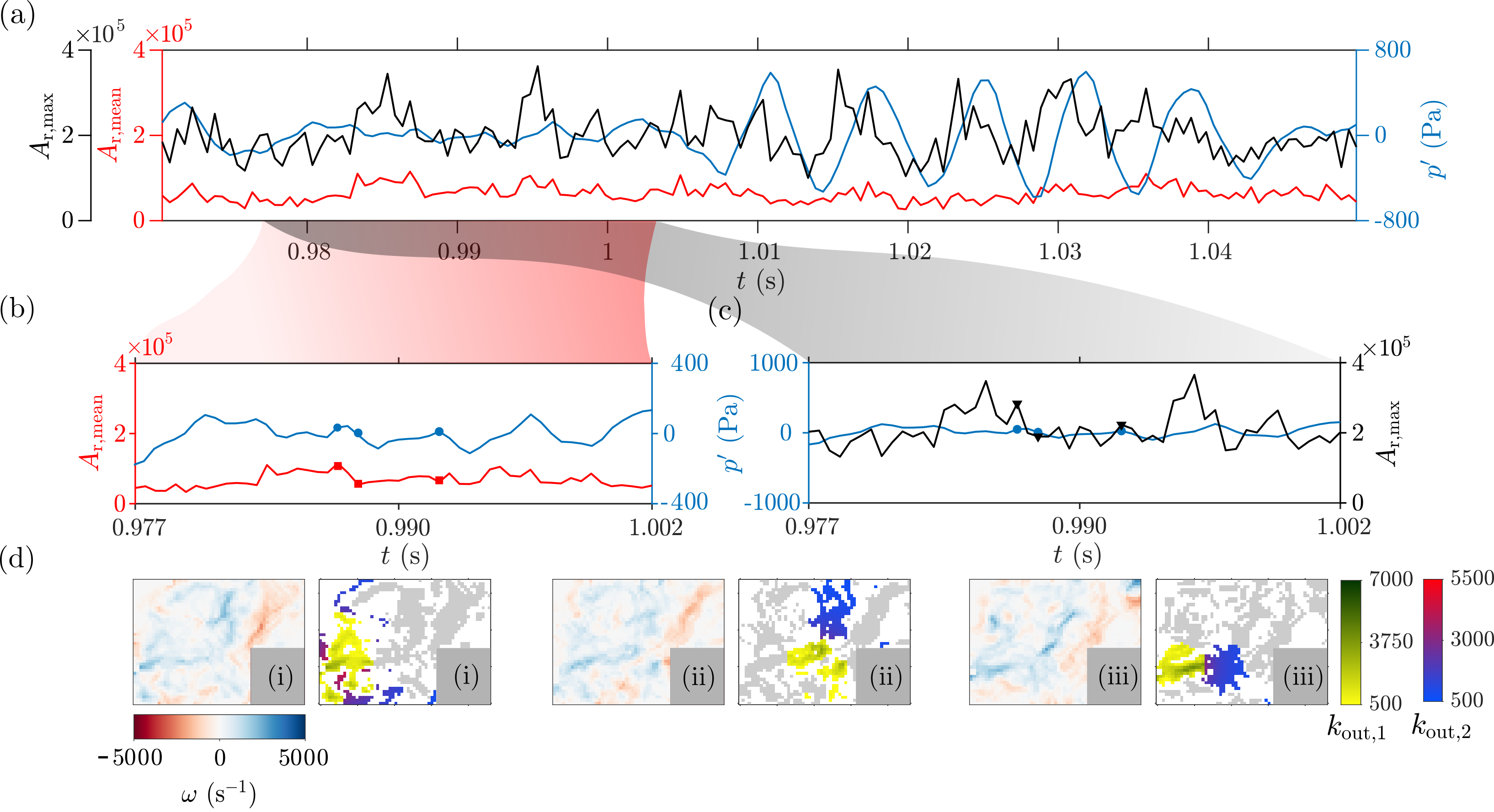}}
  \caption{(a) Time series showing the variation of $p^\prime$ fluctuations and network measures (mean $A_{\text{r,mean}}$, and maximum $A_{\text{r,max}}$ of all inter-community interactions) during an aperiodic epoch of intermittency. The zoomed-in versions are shown in (b-c). The two shaded regions indicate the portion in (a) used for the time series shown in (b-c). The vorticity field and the corresponding spatial distribution of the inter-community in- and out-strengths between the pair of communities with the maximum inter-community interaction at each of the representative time instants marked in (b-c) are shown in (d). The aperiodic nature of vortical interactions present in the reaction field causes the network measures to be aperiodic. Correspondingly, there is no specific trend in the position of the pair of communities with maximum inter-community interaction, as seen in (d).}
\label{fig_5}
\end{figure}

Similar to the aperiodic dynamics of inter-community interactions observed during the state of combustion noise, we observe aperiodicity during the aperiodic epoch of intermittency in figure \ref{fig_5}. Figure \ref{fig_5}(b-c) shows the zoomed-in versions of the time series of acoustic pressure fluctuations and network measures shown in figure \ref{fig_5}(a). Figure \ref{fig_5}(d) shows the vorticity field and the corresponding pair of maximum-interacting communities in the spatial domain. The aperiodic nature of the network measures shown in figure \ref{fig_5}(b-c) reflects the aperiodicity in the spatio-temporal evolution of vortical interactions present during the aperiodic epoch of intermittency. In figure \ref{fig_5}(d), we observe large regions of low vorticity present in the reaction field, unlike the small localized regions of high vorticity present during the state of combustion noise. Thus, compared with the fragmented nature of vortical communities present during the state of combustion noise, we observe large spatially coherent vortical communities during the aperiodic epoch of intermittency. 

Furthermore, the average of the inter-community node strengths is slightly lower than that observed during combustion noise (refer to the variation of $A_{\text{r,mean}}$ in figure \ref{fig_4}a). The lower values of $A_{\text{r,mean}}$ observed during the aperiodic epoch of intermittency are due to the large spatial extent of low vorticity regions present during the aperiodic epoch of intermittency, in contrast with the small spatial extent of high vorticity regions present during combustion noise. The difference in the values of $A_{\text{r,mean}}$ observed during the states of combustion noise and the aperiodic epoch of intermittency is an artefact of different spatio-temporal reaction fields present during the two dynamical states, even though the acoustic pressure fluctuations are of a similar order of magnitude during these two dynamical states.

\subsubsection{Periodic epoch of intermittency}

\begin{figure}
  \captionsetup{width=\textwidth}
  \centerline{\includegraphics[width=\textwidth]{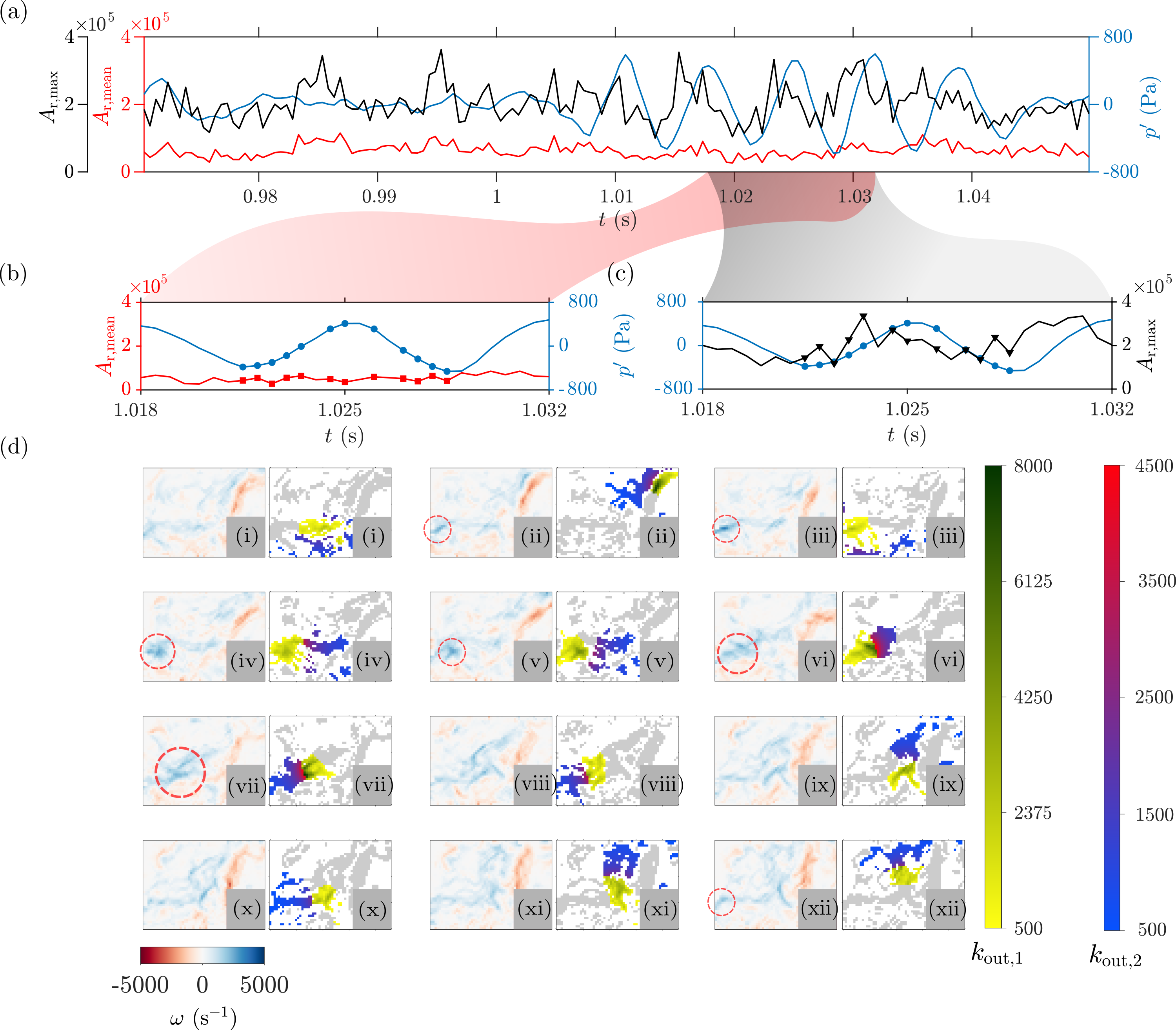}}
  \caption{(a) Time series showing the variation of $p^\prime$ fluctuations and network measures (mean $A_{\text{r,mean}}$, and maximum $A_{\text{r,max}}$ of all inter-community interactions) during a periodic epoch of intermittency. The acoustic pressure fluctuations are marked in blue. The zoomed-in versions are shown in (b-c). The two shaded regions indicate the portion in (a) used for the time series shown in (b-c). The vorticity field and the corresponding spatial distribution of the inter-community in- and out-strengths between the pair of communities with the maximum inter-community interaction at each of the representative time instants marked in (b-c) are shown in (d). The red circles in (d) indicate the downstream convection of the coherent structure shed from the backward-facing step at the time instant (ii). The location of the community with the maximum outward influence matches with the location of the downstream moving coherent structure indicated with red circles during time instants (ii-vii). Subsequently, the coherent structure gets dissipated into the reaction field. The shedding and downstream convection of the coherent structure repeats during each acoustic cycle within the periodic epoch of intermittency and is reflected in the periodic nature of $A_\text{r,max}$ shown in (c).}
\label{fig_6}
\end{figure}

During the periodic epoch of intermittency, we observe that spatially coherent regions with large vorticity magnitude begin to emerge in the reaction field (figure \ref{fig_6}). Figures \ref{fig_6}(a) and \ref{fig_6}(b,c) show the time series of acoustic pressure fluctuations and network measures and their zoomed-in versions during a periodic epoch of intermittency, respectively. Figure \ref{fig_6}(d) shows the vorticity field and the corresponding pair of maximum-interacting communities in the spatial domain. We notice that the mean of inter-community interactions $A_{\text{r,mean}}$ shows an aperiodic behaviour during the periodic epoch on intermittency. Taking the mean of all inter-community interactions possibly smears the ability of the network measure $A_{\text{r,mean}}$ to reflect the periodic spatio-temporal dynamics present in the reaction field during the periodic epoch of intermittency. However, we observe a significant correlation between the maximum of all inter-community interactions $A_{\text{r,max}}$ and the acoustic pressure fluctuations $p^\prime$ during the periodic epoch of intermittency.

One can observe the emergence of coherent structures shed from the backward-facing step during acoustic pressure minima (at time instant (ii) in figure \ref{fig_6}d). This emergence of localised coherent spatial structures is suggestive of self-organisation in the reaction field \citep{mondal2017onset, george2018pattern}. The downstream convection of the coherent structure shed from the backward-facing step is indicated through red circles in figure \ref{fig_6}(d). At the instant of the coherent structure shedding from the backward-facing step (shown at time instant (ii)), the regions of vortical communities with the maximum inter-community interaction do not match with the locations of the coherent structure shed from the backward-facing step. This is due to the coherent structure being present near the periphery of the spatial domain (adjacent to the backward-facing step), which results in low inter-community strength of the nodes belonging to the vortical communities that overlap with the recently shed coherent structure. 

Subsequently, as the coherent structure shed from the backward-facing step convects downstream (time instants ii-vii in figure \ref{fig_6}d), the inter-community interactions start to increase. Thus, the regions with vortical communities with maximum inter-community interactions match the regions of downstream convecting coherent structure until the acoustic pressure reaches the local maxima at the time instant (vii). Later, the coherent structure dissipates into the reactive flow field during the time instants marked (viii-xi). We again observe the shedding of a new coherent structure near the acoustic pressure minima at time instant (xii), and the above-discussed spatio-temporal dynamics continues for the next acoustic cycle during the periodic epoch of intermittency. This periodicity in vortical interactions that occur due to a periodic shedding and downstream convection of coherent structures during the periodic epoch of intermittency is reflected in the periodic time series of the network measure $A_{\text{r,max}}$, as seen in figure \ref{fig_6}(a).

\subsubsection{Thermoacoustic instability}

\begin{figure}
  \captionsetup{width=\textwidth}
  \centerline{\includegraphics[width=\textwidth]{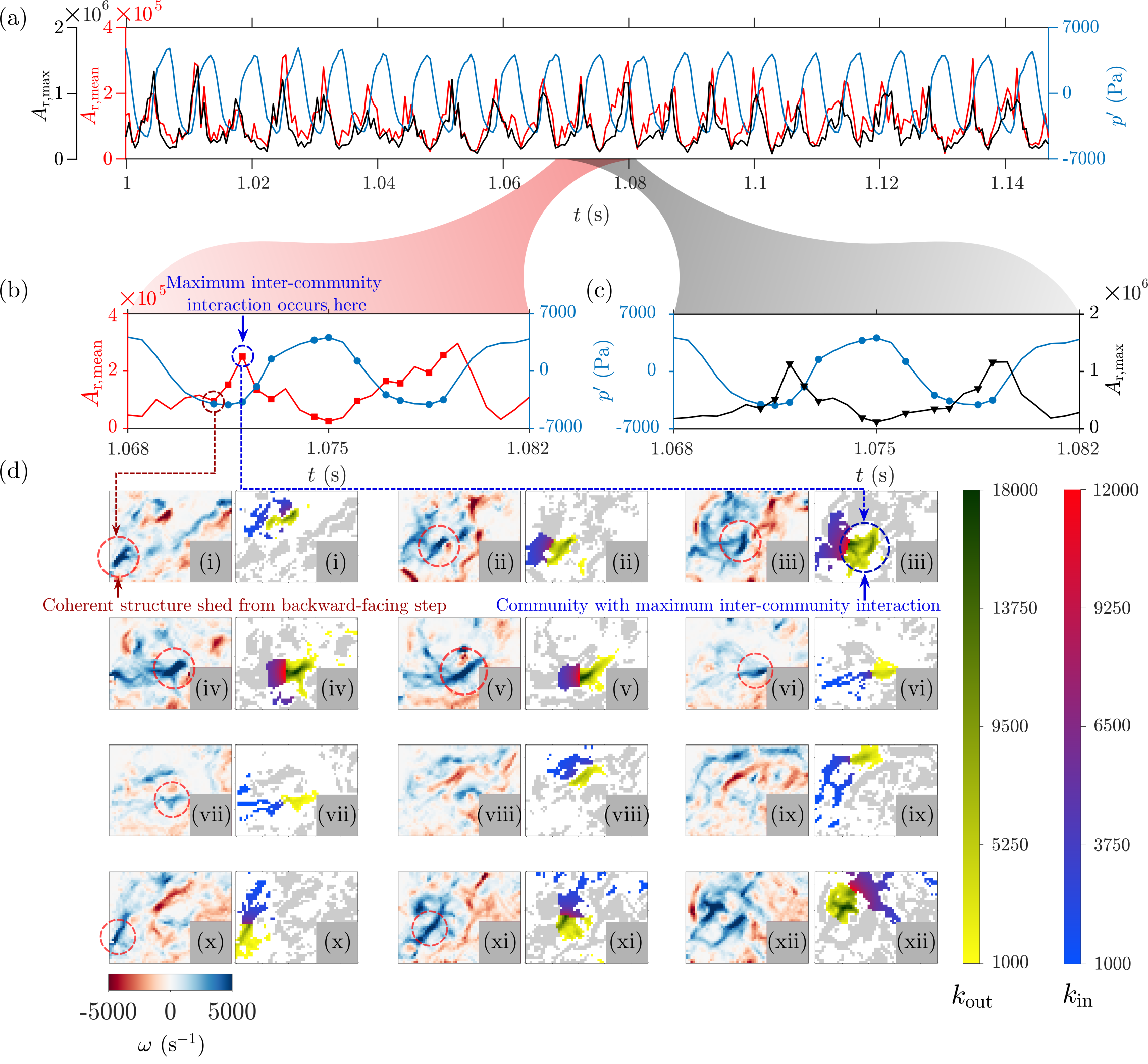}}
  \caption{(a) Time series showing the variation of $p^\prime$ fluctuations and network measures (mean $A_{\text{r,mean}}$, and maximum $A_{\text{r,max}}$ of all inter-community interactions) during the state of thermoacoustic instability. The acoustic pressure fluctuations are marked in blue. The zoomed-in versions are shown in (b-c). The two shaded regions indicate the portion in (a) used for the time series shown in (b-c). The vorticity field and the corresponding spatial distribution of the inter-community in- and out-strengths between the pair of communities with the maximum inter-community interaction at each of the representative time instants marked in (b-c) are shown in (d). The red circles in (d) indicate the downstream convection of the coherent structure shed from the backward-facing step at the time instant (i). The coherent structure sheds from the backward-facing step at the time instant marked (i) in (d). However, the maximum inter-community interaction occurs at the time instant marked (iii) in (d). This phase lag between the time instants of vortex shedding and maximum inter-community interaction is approximately $51.42 \degree$. The location of the community with the maximum outward influence matches with the location of the downstream moving coherent structure indicated with red circles during the time instants (i-vii). Subsequently, the coherent structure gets dissipated into the reaction field. The shedding and downstream convection of the coherent structure repeats itself during each acoustic cycle present within the state of thermoacoustic instability and is reflected in the periodic nature of the network measures seen in (b-c).}
\label{fig_7}
\end{figure}

Finally, in figure \ref{fig_7}, we examine the spatio-temporal dynamics of maximum-interacting vortical communities over an acoustic pressure cycle during the state of thermoacoustic instability. Figure \ref{fig_7}(a) and \ref{fig_7}(b-c) show the time series of acoustic pressure fluctuations and network measures and their zoomed-in versions, respectively. Figure \ref{fig_7}(d) shows the vorticity field and the corresponding pair of maximum-interacting communities in the spatial domain. The evolution of $A_{\text{r,mean}}$ and $A_{\text{r,max}}$ exhibits a significant periodic behaviour during the state of thermoacoustic instability. Unlike the results observed during the state of combustion noise, we notice large regions of spatially coherent vortical structures downstream of the backward-facing step during the state of thermoacoustic instability. Consequently, the vortical communities during the state of thermoacoustic instability are significantly less fragmented in the spatial domain compared with that observed during the state of combustion noise.

Notably, during the state of thermoacoustic instability, an almost consistent phase lag is present between the time instants when the coherent structures are shed from the backward-facing step (corresponding to the instants of local acoustic pressure $p^\prime$ minima) and the time instants when the vortical interactions have their maximum strength (local maxima of the network measures). This non-trivial phase lag contrasts our hypothesis that the vortical interactions will be maximum when the coherent structures are shed from the backward-facing step (during the local minima of $p^\prime$ oscillations) since the coherent structures start losing their strength as they convect downstream \citep{nair2007near,shanbhogue2009vortex}. This phase lag arises since the interactions of the shed coherent structure with other coherent structures in the flow field do not depend simply on the vorticity of the nodes constituting the coherent structure but also on the distance of the coherent structure from other coherent structures. Although the coherent structure at the time of its shedding from the backward-facing step exhibits the most significant vorticity strength values, its considerable distance from other coherent structures in the flow field contributes to low vortical interactions. A decrease in this distance contributes to an increase in the magnitude of vortical interactions as the coherent structure convects downstream towards the middle of the flow field. Subsequently, the loss of vorticity strength of the coherent structure dominates the small distance among different coherent structures, and the magnitude of vortical interactions decreases.

We can observe the above-mentioned behaviour by examining the spatio-temporal dynamics of the shedding of the coherent structure from the backward-facing step during the acoustic pressure minima (at time instant (i)) during the state of thermoacoustic instability. The downstream convection of the coherent structure shed from the backward-facing step at the minima of $p^\prime$ oscillations is indicated through red circles in figure \ref{fig_7}(d). The regions of vortical communities with the maximum inter-community interaction match with the regions of coherent structure shed from the backward-facing step during the time instants (ii-vii). This matching of spatial regions of coherent structure shed from the backward-facing step and the community with maximum outward interaction is clearly visible till the acoustic pressure reaches the local maxima at the time instant (vii). Subsequently, as the coherent structure dissipates into the reactive flow field and the magnitude of vorticity present in the reactive field decreases during the time instants marked (viii-ix), it is difficult to discern the downstream movement of the coherent structure being discussed. We again observe the shedding of a new coherent structure near the acoustic pressure minima at time instant (x), and the above-discussed spatio-temporal dynamics continues for the next acoustic cycle.

\begin{figure}
  \captionsetup{width=\textwidth}
  \centerline{\includegraphics[width=\textwidth]{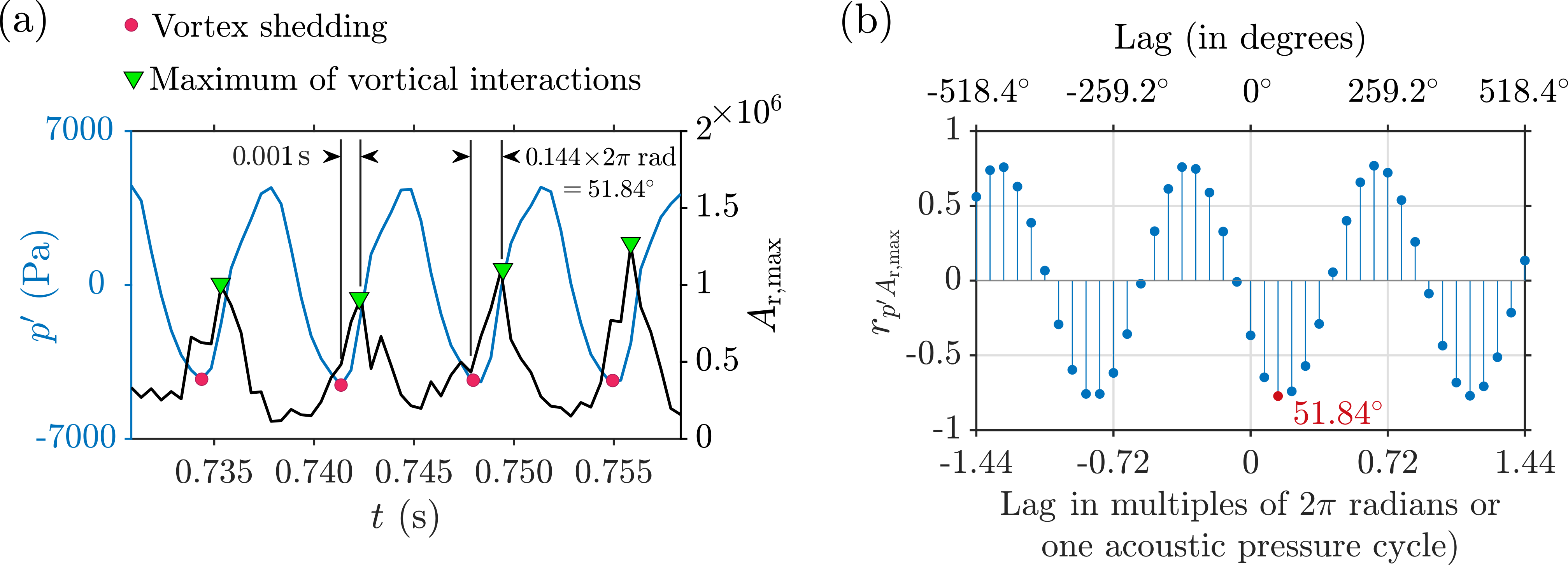}}
  \caption{(a) Time series of $p^\prime$ fluctuations and $A_\text{r,max}$, indicating the instants of vortex shedding and maximum vortical interactions through red circles and green triangles, respectively. (b) Shows the cross-correlation between $p^\prime$ and $A_\text{r,max}$, denoted by $r_{p^\prime A_\text{r,max}}$ and the corresponding lags in terms of multiples of acoustic pressure cycles ($2\pi$ rad is equivalent to one acoustic pressure cycle) and degrees. The minima of $r_{p^\prime A_\text{r,max}}$ is at $52 \degree$, indicating the lag between the occurrences of $p^\prime$ minima (i.e., vortex shedding from the backward-facing step) and maximum of vortical interactions.}
\label{fig_8}
\end{figure}    

As mentioned above, one can expect the vortical interactions in the combustor to be maximum at the instant of shedding as the shed vortices lose their strength via dissipation to the surroundings during their convection downstream. However, we observe a lag between the occurrences of vortex shedding and maximum vortical interactions in figure \ref{fig_7}(a). In figure \ref{fig_7}(b), this phase lag can be observed between the time instants marked (i) and (iii) in figure \ref{fig_7}(d). We notice the newly formed coherent structure shedding from the backward-facing step at the acoustic pressure minima marked 1 in figure \ref{fig_7}(b). However, the maximum inter-community interaction occurs at the time instant marked 3. To calculate the phase difference between points (i) and (iii), we consider the fact that there are 14 to 16 points with velocity fields obtained from 2C PIV measurements in an acoustic pressure cycle during the state of thermoacoustic instability. Thus, the phase lag between two adjacent 2C PIV data points varies between $360\degree / (16 - 1) = 24\degree$ and $360\degree / (14 - 1) = 27.69\degree$. Since the time instants of vortex shedding and maximum inter-community interaction occur at points marked 1 and 3 in figure \ref{fig_7}(b), the phase lag between these two points is approximately in the range of $24\degree \times 2 = 48\degree$ to $27.69\degree \times 2 = 55.38\degree$.

To perform the above phase lag analysis for the entire time series of $p^\prime$ and $A_\text{r,max}$ corresponding to the state of thermoacoustic instability, we use cross-correlation between the time series of $p^\prime$ fluctuations and $A_\text{r,max}$ to compute the phase lag between the vortex shedding and the occurrence of maximum vortical interactions. The minima of $p^\prime$ fluctuations correspond to the instants of vortex shedding. A few representative instants of phase lags and the cross-correlation ($r_{p^\prime A_\text{r,max}}$) are shown in figures \ref{fig_8}(a) and \ref{fig_8}(b), respectively. The minima of cross-correlation closest to $0 \degree$ in figure \ref{fig_8}(b) indicates the phase lag where $p^\prime$ and $A_\text{r,max}$ are maximum negatively correlated. We consider maximum negative correlation since the minima of $p^\prime$ fluctuations and maxima of $A_\text{r,max}$ are used for computing the cross-correlation. We observe that a significant phase lag of about $51 \degree$ is present between the instants of vortex shedding from the backward-facing step and the maximum inter-community interactions.

The discrepancy between the phase-averaged regions of vorticity and participation coefficient, as discussed in Section \ref{sec:phase_averaged}, can now be understood by interpreting the phase lag mentioned in the previous paragraph. The region of maximum outward inter-community interaction is located between the backward-facing step and the centre of the flow field. Considering the phase angle, where $0 \degree$ marks the instant a vortex is shed at the upstream backward-facing step (cf. figure \ref{fig_3}a), an increase in phase angle corresponds to the downstream convection of the shed vortex, as shown in figure \ref{fig_3}(c). Furthermore, figure \ref{fig_3}(b) shows that the region of maximum outward inter-community interaction corresponds to a phase angle of approximately $45 \degree$. This is very close to the phase lag of $52 \degree$ mentioned in the previous paragraph. Thus, even as the vortex shed from the backward-facing step continues to move downstream, the region of maximum outward inter-community interaction remains around $52 \degree$, explaining the observed discrepancy between the phase-averaged regions of vorticity and participation coefficient. This shows that in addition to the crucial role coherent structures play in combustion systems, the interaction of vortical elements is also necessary to comprehensively understand the flow physics in combustion systems.

\subsection{Insights into critical region detected for open-loop control}

Having analyzed the spatio-temporal evolution of maximum-interacting pairs of communities during different dynamical states exhibited by the turbulent combustor, we now focus on the relevance of maximum inter-community interactions toward mitigating thermoacoustic instability. The turbulent combustor being discussed in the current study exhibits vortex-driven thermoacoustic instability. Thus, the mitigation of thermoacoustic instability in this type of combustor is intricately related to leveraging our understanding of the vortex-driven spatio-temporal dynamics. Open-loop control strategies are widely used in practical combustion systems due to their minimal maintenance requirements and high durability. One of the open-loop control methods is to target certain \textit{critical} regions using secondary fuel or air to perturb the flow regions responsible for controlling the overall dynamics of the system. \citet{ghoniem2005stability} and co-workers \citep{altay2007impact, altay2010mitigation} determined the optimum region as the region of ﬂame anchoring and considered a steady injection of secondary air for achieving control in a dump combustor. \citet{uhm2005low} considered the region of local maxima in the heat release rate field to be the critical region. In a similar study, \citet{tachibana2007active} optimized the secondary fuel injector parameters using the Rayleigh Index distribution to control thermoacoustic instability in a swirl-stabilized lean premixed combustor actively.

\begin{figure}
  \captionsetup{width=\textwidth}
  \centerline{\includegraphics[width=\textwidth]{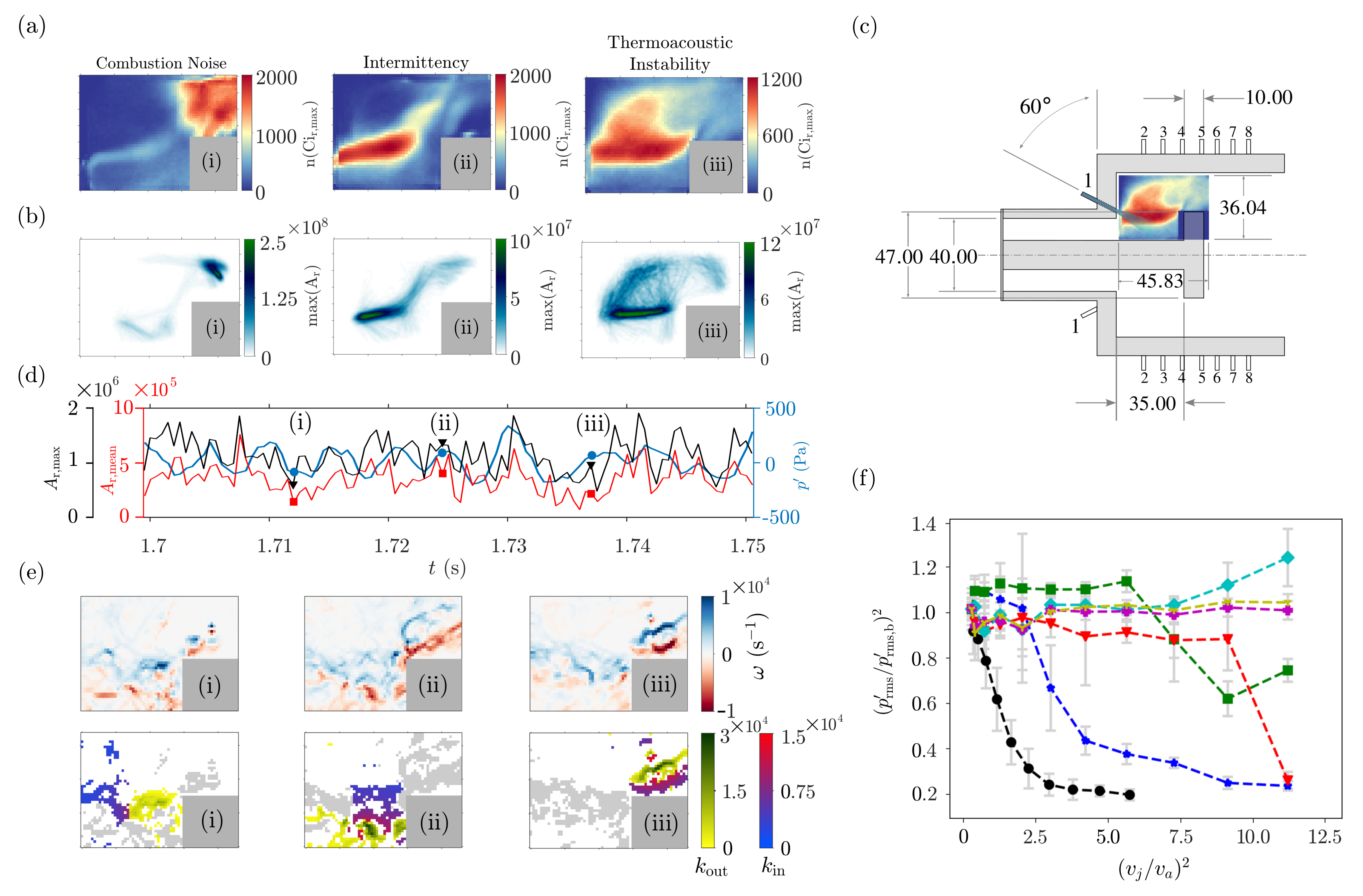}}
  \caption{(a) The distribution of the spatial location of vortical communities with the maximum inter-community interactions during the states of combustion noise, intermittency, and thermoacoustic instability. Here, n(Ci$_{\text{r,max}}$) denotes the number of overlapped vortical communities used to plot the distribution of the spatial location of vortical communities with the maximum inter-community interactions. (b) Plot of kernel smoothed intensity function of the maximum inter-community interactions. Consistent with the results discussed in figure \ref{fig_4}(d), most of the communities with maximum inter-community interactions are present above the bluff body during the state of combustion noise. (c) Shows the cross-section of the combustion chamber showing the microjet injector locations. (d) Shows the aperiodic network measures and the $p^\prime$ fluctuations during the state of suppression of thermoacoustic instability when air microjet is injected through injector number 1. (e) Shows the spatio-temporal variation of vorticity and regions of maximum inter-community interaction during the state of suppression. (f) Is adapted with permission from \citet{krishnan_2021} and shows the variation of normalized root-mean-square values of $p^\prime$ due to air microjet injection at different locations shown in (c). The horizontal axis shows the total momentum flux ratio of the air microjet. The vertical error bars denote the standard deviation. The air microjet injected via port 1 induces the most efficient suppression of thermoacoustic instability with the least momentum flux ratio of the air microjet. (c) Illustrates how air microjet injection via port 1 appears to disrupt the region with maximum inter-community interactions during the state of thermoacoustic instability. While the disruption may not be significant, some degree of correlation can be observed between disruption of the region of maximum inter-community interaction and suppression of thermoacoustic instability.}
\label{fig_9}
\end{figure}

The first step toward suppressing thermoacoustic instability is to heuristically find the critical region where the secondary injection to achieve open-loop control should be targeted. The high correlation between network measures and acoustic pressure fluctuations during the state of thermoacoustic instability, as seen in figure \ref{fig_7}, motivates us to conjecture that communities with maximum inter-community interactions play a critical role towards the macroscopic dynamics exhibited by the combustor. Disrupting the interactions between the communities having the maximum interaction during thermoacoustic instability can help us to break the positive coupling between the vortical interactions and acoustics in the combustor, thus suppressing the high-amplitude acoustic pressure fluctuations. This would, in turn, correspond to perturbing the closely positioned regions of high vorticity that are detected as the largest interacting pair of communities in the network space. While the results suggest that targeting these regions may help disrupt the positive coupling between vortical interactions and acoustics in the combustor, it is important to note that further rigorous causal analysis can be carried out to quantitatively establish a correlation between the injector locations and the suppression of thermoacoustic instability

Thus, our final objective is to understand the occurrence of the critical regions in the reaction flow field through the lens of vortical interactions analyzed via complex networks. To this end, we plot the spatial distribution of vortical communities having the maximum interaction (at all time instants) during the states of combustion noise, intermittency, and thermoacoustic instability in figure \ref{fig_9}a(i-iii). For each dynamical state, 3796 snapshots (equivalent to 300 acoustic cycles present during the state of thermoacoustic instability) are utilized for this analysis. The spatial distribution of the communities having maximum interaction is plotted by first detecting the pair of such communities at each time instant during a dynamical state. All such pairs of communities having the maximum inter-community interactions at all the time instants are subsequently overlapped in the spatial domain to identify the region where such influential communities are most likely to be found. The colorbar for figure \ref{fig_9}(a) represents the number of vortical communities with maximum inter-community interactions detected during the three different dynamical states. We also plot the spatial distribution of weights of the maximum inter-community interactions in figure \ref{fig_9}(b), where we plot the corresponding kernel smoothed intensity function of $A_{\text{r,max}}$. Here, $A_{\text{r,max}}$ is the magnitude of the largest inter-community interaction and is equal to the largest element of the reduced adjacency matrix. 

In figure \ref{fig_9}a-b(i), we observe that the influential communities are present primarily above the bluff body during the state of combustion noise. These vortical communities correspond to the small-sized clockwise vortices shed from the tip of the bluff body due to the flow turning at the corner of the bluff body. We also detect a distinct region with maximum-interacting vortical communities downstream of the backward-facing step during the state of intermittency (figure \ref{fig_9}a-b(ii)). The periodic and the aperiodic epochs are considered together while calculating the region of the maximum-interacting pair of interactions during the state of intermittency. The vortices are shed almost periodically at the backward-facing step for short epochs of time during the periodic epoch of intermittency, after which they may break down into smaller vortices or impinge on the walls of the combustor or bluff body. The location of this region is consistent with the critical regions detected during the state of intermittency by plotting quantities such as Hurst exponent and the node strengths of the $p^\prime \dot{q}^\prime$-$\omega$ multilayer network, as seen in \citet{roy2021critical} and \citet{tandon2022multilayer}, respectively. Other physical measures such as the Fourier amplitude of turbulent velocity ﬂuctuations \citep{roy2021critical}, weighted closeness centrality in a weighted spatial correlation network obtained from the velocity field in \citet{krishnan2019mitigation}, and node strength of the phase-averaged vortical network \citep{krishnan_2021} can identify critical region only during the state of thermoacoustic instability, not during intermittency. 

Finally, during thermoacoustic instability, the region with maximum-interacting vortical communities increases in size and extends from the backward-facing step to the upstream boundary of the bluff body, as seen in figure \ref{fig_9}a-b(iii). A significant part of this critical region, calculated through the present methodology during the state of thermoacoustic instability, is consistent with the corresponding critical regions detected through the spatial distribution of Hurst exponent (cf. figure 9(c) in \citep{roy2021critical}), weighted closeness centrality of weighted spatial correlation networks constructed from velocity field (cf. figure 1(c) in \citep{krishnan2019mitigation}), and node strength of the phase averaged vortical network (cf. figure 13(b) in \citep{krishnan_2021}). The consistency of the region having maximum-interacting vortical communities detected in the current study with that of critical regions detected in previous studies provides strong evidence that the community-based formulation can effectively capture critical regions in reactive flow systems and shed light on the spatio-temporal dynamics of vortex-driven thermoacoustic systems.

Open-loop control requires knowledge about the relative importance of different regions in the ﬂow ﬁeld. In the experimental set-up used in the present study, \citet{krishnan_2021} introduced microjets of air into the critical region to suppress thermoacoustic instability. Microjets of air steadily injected through the injector numbered 1 (cf. figure \ref{fig_9}c) resulted in the most efficient mitigation of thermoacoustic instability compared with injection via other ports. We conjecture that this targeted control disrupted the key vortical communities with the maximum inter-community interactions present during the state of thermoacoustic instability, thus mitigating instability. During the state of suppression of thermoacoustic instability through microjet injection via injector 1, we observe in figure \ref{fig_9}(d) that the time series of the mean ($A_{\text{r,mean}}$) and maximum ($A_{\text{r,max}}$) of all inter-community interactions exhibit an aperiodic dynamics, similar to that present in the time series of acoustic pressure fluctuations. The aperiodic dynamics exhibited by the network measures is similar to that observed during the state of combustion noise shown in figure \ref{fig_4}(a). Figure \ref{fig_9}(e) shows the snapshots of vorticity distribution and inter-community node strengths of the pair of communities with the largest inter-community interaction. Amongst all the air microjet injection ports shown in figure \ref{fig_9}c), injecting air through port 1 leads to the maximum suppression of thermoacoustic instability (figure \ref{fig_9}f). 

While the data presented in figure \ref{fig_9}(f) demonstrates a correlation between microjet injector placement and suppression efficiency, it should be noted that the current analysis does not quantitatively establish the optimality of this placement through a causal investigation. The observed suppression bears a noticeable correlation with the disruption of key vortical communities identified through vortical network analysis. However, further studies involving a more comprehensive quantitative approach are required to investigate these findings. For instance, a network dynamical system could be developed by examining the impact of perturbing specific spatial locations (network nodes) using microjets of air, with the perturbations modelled as varying magnitudes of external disturbances applied to the network.

\section{Concluding remarks}\label{sec:conclusion}

A great deal of previous research into thermoacoustic instability has focused on the role of coherent structures in driving thermoacoustic instability in turbulent combustors. \citet{rogers1956mechanism} provided the earliest description of the role of vortices in thermoacoustic instability. They showed that a transverse acoustic mode excited in an acoustic duct triggered the shedding of vortices in the reaction field. Subsequently, a significant portion of the work done since the 1980s has extended our knowledge of understanding how coherent structures contribute to thermoacoustic instability \citep{zukoski1985combustion, poinsot1987vortex, hussain2022simulations}. 

However, several questions remain unanswered on the dynamics of the interaction of vortical structures in turbulent thermoacoustic systems. This paper argues that, in contrast to the hypothesis that vortical interactions will be close to the instants of shedding of coherent structure \citep{nair2007near,shanbhogue2009vortex}, these interactions attain their maxima downstream in the flow field. To show this, we compute the critical interactions between regions of intense vorticity through induced velocities calculated using the Biot-Savart Law. These induced velocities are used to construct time-varying weighted directed networks, termed vortical networks. The vortical networks are used to identify the communities or modular structures in the unsteady flows through the modularity maximization algorithm. The temporal variation of the vortical interactions is captured via the time series of mean and maximum inter-community interactions. Subsequently, we show a consistent phase lag between the instants of shedding of coherent structures during the state of thermoacoustic instability and the instants where the vortical interactions achieve their maximum value. The strong delayed correlations observed between the statistics of vortical community interactions and the acoustic pressure fluctuations during the state of thermoacoustic instability emboldens us to conjecture that the key vortical community interactions play a critical role in the behaviour exhibited by the combustor. Taking a cue from this line of thought, we also show that the regions exhibiting the largest inter-community interactions are consistent with the critical regions detected during the states of intermittency and thermoacoustic instability in previous studies. Perturbing the regions of the most influential pair of communities disrupts the largest inter-community interactions in a temporally averaged manner and leads to the maximum suppression of thermoacoustic instability. Our results complement the findings in earlier works by \citet{krishnan2019mitigation}, \citet{roy2021critical}, \citet{krishnan_2021}, \citet{tandon2022multilayer}, and \citet{premchand2023control} on the same experimental setup, thus validating our approach.

While acknowledging the limitations inherent in our study due to the discrepancy between the actual three-dimensional flow field and the two-dimensional data acquired, we are hopeful that our work effectively elucidates the relevant physics of the complex flow field of a turbulent thermoacoustic system. Our optimism stems from the fact that the significant features of the turbulent flow field, such as the periodic shedding of vortices during the states of periodic epochs of intermittency and thermoacoustic instability, are adequately captured in the two-dimensional PIV data. Moreover, the two-dimensional PIV data captures irregular spatio-temporal behaviour of the flow field during the states of combustion noise and aperiodic epochs of intermittency, consistent with our understanding of these dynamical states. Nevertheless, we recognize that three-dimensional simulations or three-component tomographic PIV can potentially reveal additional information that our current two-dimensional PIV data cannot fully capture. Future studies incorporating three-dimensional experimental and computational data can complement and extend our findings, providing a more comprehensive understanding of vortical interactions in turbulent reactive flow fields. Furthermore, baroclinic torque and dilatation have been proven to play a pivotal role in the vorticity dynamics of reactive flow fields \citep{geikie2018pressure, geikie2021turbulent, rising2021effects}. We look forward to future work utilizing three-dimensional stereoscopic PIV and Mie scattering data, acquired at a sufficiently high sampling rate, to investigate the role of different vorticity generating mechanisms in the dynamics of vortical interactions in turbulent combustors.

In closing, the current method has the potential for widespread application in combustion and other fluid-mechanical systems. First, it enables us to recognize and quantify how the interactions between different vortical communities correlate with the macroscopic dynamics exhibited by the fluid-mechanical systems. In addition to the already well-established role of coherent structures in combustion systems, analyzing the interaction between different vortical elements is essential to understanding the spatio-temporal dynamics of combustion systems. Secondly, for combustion systems susceptible to thermoacoustic instability, the distribution of vital vortical communities during the state of intermittency can be used to explain the occurrence of critical regions that have been shown to drive thermoacoustic instability. Thirdly, the present work provides a potential way forward to develop low-order models for a turbulent thermoacoustic system using the statistical behaviour of communities, the inter- and intra-community strengths, and the locations of communities in the reactive field of the combustor. Finally, the present methodology provides a path for the future development of advanced flow control efforts to alter the dynamics of vortical structures in turbulent fluid-mechanical systems. The present methodology is generalizable to arbitrary thermofluid systems where instability control is crucial.

\section*{Acknowledgments}
We thank Jayesh M.\ Dhadphale, Samarjeet Singh, and Gaurav Chopra for countless fruitful discussions. We acknowledge Thilagaraj S., Anand Selvam, Manikandan Raghunathan, and Abin Krishnan for providing the experimental data. Ankit Sahay is grateful to the Ministry of Education for the Half-Time Research Assistantship (HTRA). The authors are grateful to Prof.\ Kunihiko Taira for fruitful discussions.

R.\ I.\ Sujith is grateful for the funding from the Office of Naval Research Global (Grant No.\ N62909-18-1-2061; Funder ID: 10.13039/100007297), and the Institute of Eminence (IoE) initiative of IIT Madras (SB/2021/0845/AE/MHRD/002696). This research used resources of the Oak Ridge Leadership Computing Facility at the Oak Ridge National Laboratory, which is supported by the Office of Science of the US Department of Energy under Contract No. DE-AC05-00OR22725.

\appendix
\section{Vorticity thresholding}\label{AppendixA}

\begin{figure}
  \captionsetup{width=1\textwidth}
  \centerline{\includegraphics[width=\textwidth]{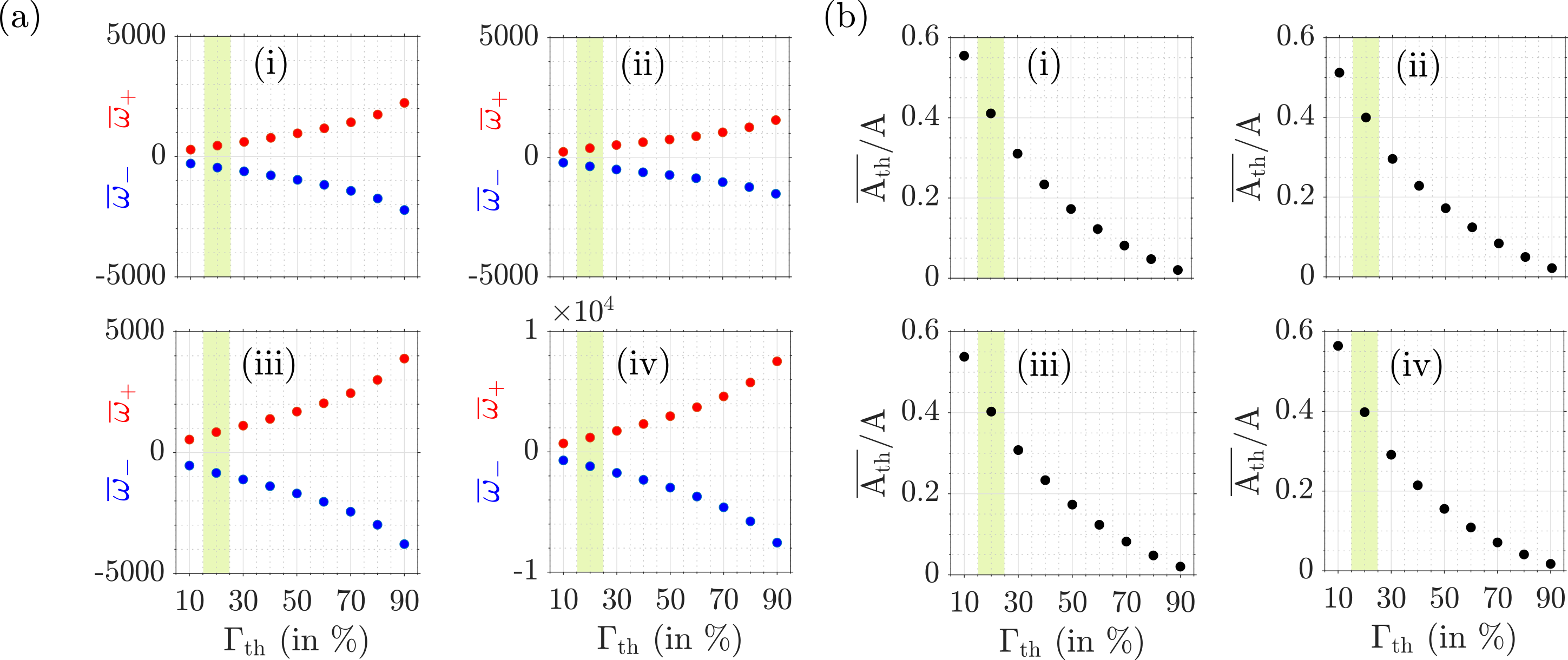}}
  \caption{(a) Variation of mean positive ($\overline{\omega}_{+}$) and mean negative ($\overline{\omega}_{-}$) vorticity threshold values with respect to the total circulation threshold ($\Gamma_{th}$ in \%) for the states of (i) combustion noise, (ii) intermittency, (iii) thermoacoustic instability, and (iv) suppression through air microjet injection. The fluid elements with vorticity $\omega \in [\overline{\omega}_{-},\overline{\omega}_{+}]$ are disregarded to exclude the bottom $\Gamma_{th}$  \% of the total circulation present in the reaction field, thus capturing the influential regions of the flow. The $\overline{\omega}_{+}$ and $\overline{\omega}_{-}$ values present on the green band are subsequently used as thresholds while using the Louvain algorithm to detect vortical communities. (b) Variation of the mean area percentage of the reaction field $\overline{A}_\text{th}$ occupied by the vortical communities. We do not consider the area of the bluff body mask while calculating the percentage with respect to the total area of the reaction field. For all the cases, the resolution parameter $\gamma$ used in the Louvain algorithm is kept consistent at 1 \citep{meena2018network, meena2021identifying}.}
\label{fig_app1}
\end{figure}

We use vorticity thresholding to utilise fluid elements with sufficiently high vorticity magnitudes for vortical community detection. For each time instant, we calculate the range of vorticity values $\omega \in [\omega_{-},\omega_{+}]$ that contribute to a lower range of 10\% of the total circulation $\Gamma$ present in the reaction field at that time instant (for instance, $\Gamma_{th} = 10 \% = 0.1 \times \Gamma$). Finally, we compute the average of all the vorticity thresholds calculated for this circulation threshold, i.e., $\overline{\omega}_{-} = \dfrac{1}{N} \sum \omega_{-}$ and, $\overline{\omega}_{+} = \dfrac{1}{N} \sum \omega_{+}$. We repeat the above procedure removing different lower ranges of the total circulation, i.e., 20\%, 30\%, $\ldots$, 90\%. We plot the corresponding vorticity threshold values in figure\ref{fig_app1}(a). Here, $N = 3796$ is the total number of snapshots of temporal vorticity fields available for each dynamical state. 

The fluid elements with vorticity outside the aforementioned interval, i.e., the fluid elements with $\omega \notin [\omega_{-},\omega_{+}]$, are subsequently used to calculate the vortical communities at each time step. The area occupied by all the vortical communities for a circulation threshold $\Gamma_{th}$ is denoted by $A_{th}$. The average of $A_{th}$ calculated at all the time instants, i.e., $\overline{A}_{th} = \dfrac{1}{N} \sum A_{th}$, is plotted in figure \ref{fig_app1}(b) as a fraction of the total area of the spatial domain. 

The vorticity values corresponding to the lower range of 20\% of the total circulation are used for vorticity thresholding in the current study. The threshold values are highlighted with green background in figure \ref{fig_app1}(a), and are equal to $|\omega|$ $=$ $456$ $s^{-1}$, $380$ $s^{-1}$, $843$ $s^{-1}$, and $532$ $s^{-1}$  during the states of combustion noise, intermittency, thermoacoustic instability, and microjet injection, respectively. The robustness of the thresholds is evaluated by noting that the qualitative results do not change significantly even if the threshold values are modified in $\pm 50$ $s^{-1}$ range around the values considered in the current study. Approximately 40\% of the area of the reaction field considered in the current study is occupied by the detected vortical communities during all the dynamical states.  

\section{The Louvain algorithm and its non-deterministic nature}\label{AppendixB}

\begin{figure}
  \captionsetup{width=1\textwidth}
  \centerline{\includegraphics[width=1\textwidth]{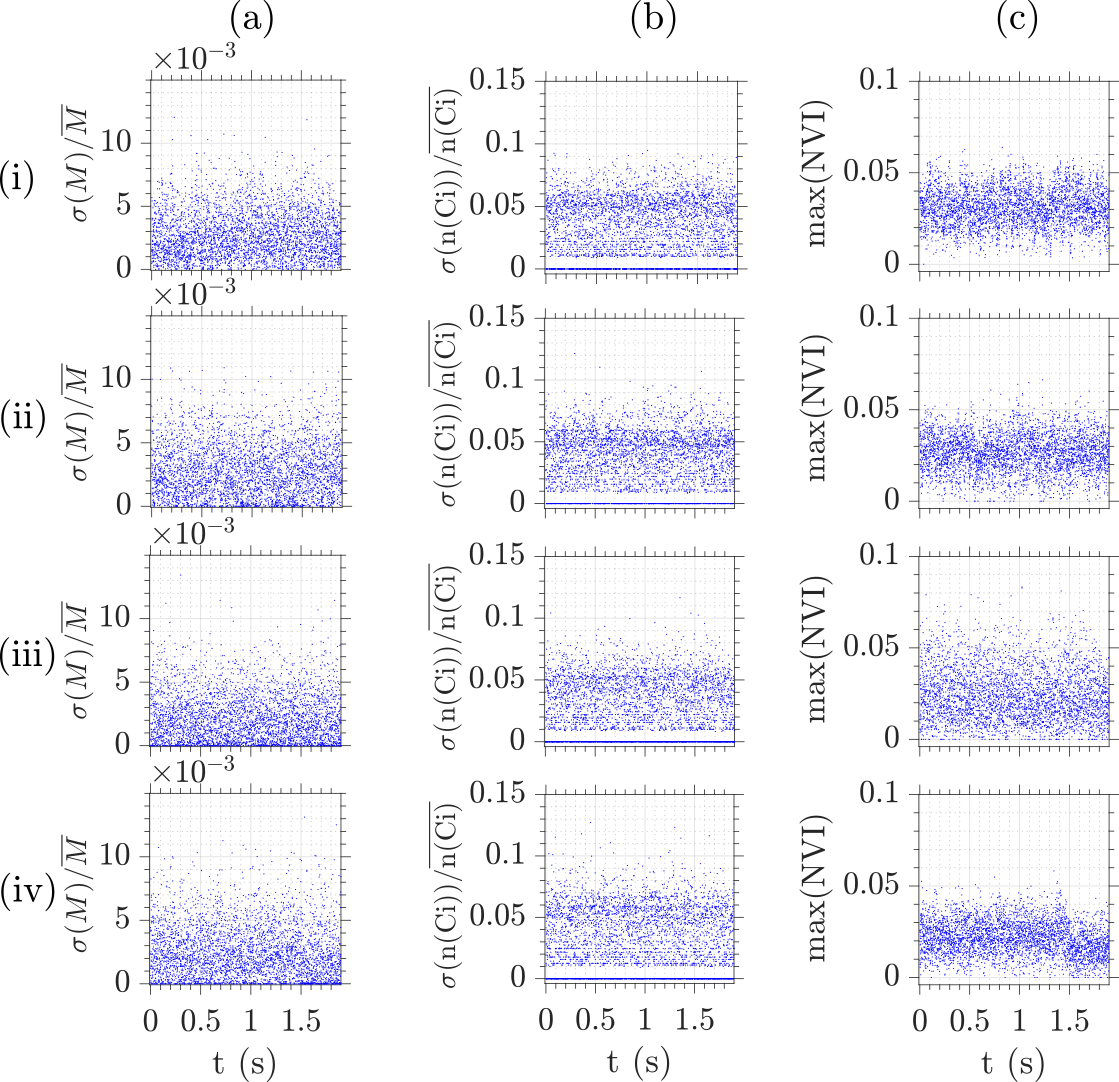}}
  \caption{(a) Plot of the standard deviation of modularity values, (b) coefficient of variation of the number of communities, and (c) maximum normalized variation of information calculated through 100 implementations of the Louvain algorithm at each time instant during the states of (i) combustion noise, (ii) intermittency, (iii) thermoacoustic instability, and (iv) suppression of thermoacoustic instability through air microjet injection.}
\label{fig_app2}
\end{figure}

Community detection algorithms are widely employed to bring out the underlying mesoscopic structure, thereby revealing the clustered nature of complex networks. While there is no formal definition for the concept of communities, the idea is to group nodes into distinct, densely connected parts, with the density between parts being low. Researchers continue to develop scalable methods to reveal densely connected groups due to the constant rise in data and dimensions. Due to the efficiency of their implementation, algorithms that maximize modularity have garnered considerable attention. \cite{newman2004fast} created modularity as a quality function to assess the quality of a partition as a community structure. It compares the existence of an edge within a community with the probability of having such an edge between corresponding vertices in a random network with the same degree distribution. While it is known that modularity has some limitations \citep{fortunato2007resolution, lancichinetti2011limits}, 
heuristic-based algorithms that maximize this measure remain the most effective on large networks to this day \citep{blondel2008fast, leicht2008community}. The Louvain algorithm is one of the most popular algorithms used to maximize modularity \citep{blondel2008fast}. 

The Louvain algorithm begins by creating communities of size one, where each node in the network forms a community. The algorithm then proceeds by executing two stages. The algorithm assigns a node $i$ to a community of a neighbour $j$ to increase the partition's modularity in the first stage. The first stage is repeated until the modularity cannot be increased. This procedure generates an initial network partition. In the second stage of the Louvain algorithm, each partition's community is considered a supernode. If at least one edge exists between nodes of each community that the supernode represents, then the supernodes are connected. After this second stage is completed, the algorithm iterates until the modularity can no longer increase.

The Louvain algorithm is non-deterministic due to the changes in the following two steps between different runs: (i) the order in which all the nodes are processed and (ii) the order in which two pairs of nodes are assigned to two different communities if the increase in modularity is equal for both the assignments \citep{cordeiro2016dynamic}. Thus, the community affiliations may vary between different implementations of the Louvain algorithm even when the exact same data set is used. 

To evaluate the extent of consistency between community structures obtained from different implementations of the Louvain algorithm on the same vortical network, we implement the Louvain algorithm 100 times at each time instant and calculate the following quantities for all the vortical networks corresponding to all the time instants: 

\begin{enumerate}
    \item Coefficient of variation of modularity, $\sigma(M)/\overline{M}$, obtained from a hundred different community affiliations at each time instant. The coefficient of variation is the ratio of the standard deviation to the mean of the modularity values obtained from different runs of the Louvain algorithm and shows the extent of variability with respect to the mean of the modularity values \citep{castagliola2013monitoring}. The higher the coefficient of variation is, the greater the difference between the modularity values obtained between different implementations of the Louvain algorithm on the same vortical network.
    
    \item Coefficient of variation of the number of communities obtained from 100 different community affiliations at each time instant.
    
    \item Maximum of normalized variation of information (NVI) to indicate the maximum deviation between any two community affiliations out of 100 different implementations of the Louvain algorithm for each time instant \citep{meilua2007comparing, karrer2008robustness}. The NVI between two community affiliations $X$ and $Y$ is given by
    \begin{equation}
        NVI(X, Y) = - \sum_{ij} r_{ij} \left[\text{log} \left( \dfrac{r_{ij}}{p_i}\right) + \text{log} \left( \dfrac{r_{ij}}{q_j}\right) \right], 
    \end{equation}
    where,
  \begin{align*} 
     p_i = |X_i|/n, q_j = |Y_j|/n, r_{ij} = |X_i \cap Y_j|/n, n = \sum_{i}|X_i| = \sum_{j}|Y_j|.
  \end{align*}
    An NVI equal to one indicates the maximum difference between two clustering of the same complex network, which occurs when each node is assigned to its own community in one clustering, and the entire network is assigned to one single community in the other clustering.
\end{enumerate}

In figure \ref{fig_app2}, we observe that the three aforementioned quantities calculated are quite low for all four dynamical states, indicating that different implementations of the Louvain algorithm give us statistically similar community affiliations for a dataset.

\section{Spatial coherence of vortical communities}\label{AppendixC}

In figure \ref{fig_5},  we observe that the communities detected during the state of combustion noise are highly fragmented in comparison with the communities detected during the other two dynamical states of intermittency and thermoacoustic instability. The highly fragmented communities present during the state of combustion noise are associated with the chaotic spatio-temporal dynamics observed in the reaction field. Furthermore, most of the pairs of communities with the maximum inter-community interactions are present above the bluff body and their spatial locations match with the coherent structures shed from the upstream tip of the bluff body. On the other hand, most of the highly interacting communities observed during the states of intermittency and thermoacoustic instability match with the locations of coherent structures shed from the backward-facing step.

\begin{figure}
  \captionsetup{width=1\textwidth}
  \centerline{\includegraphics[width=1\textwidth]{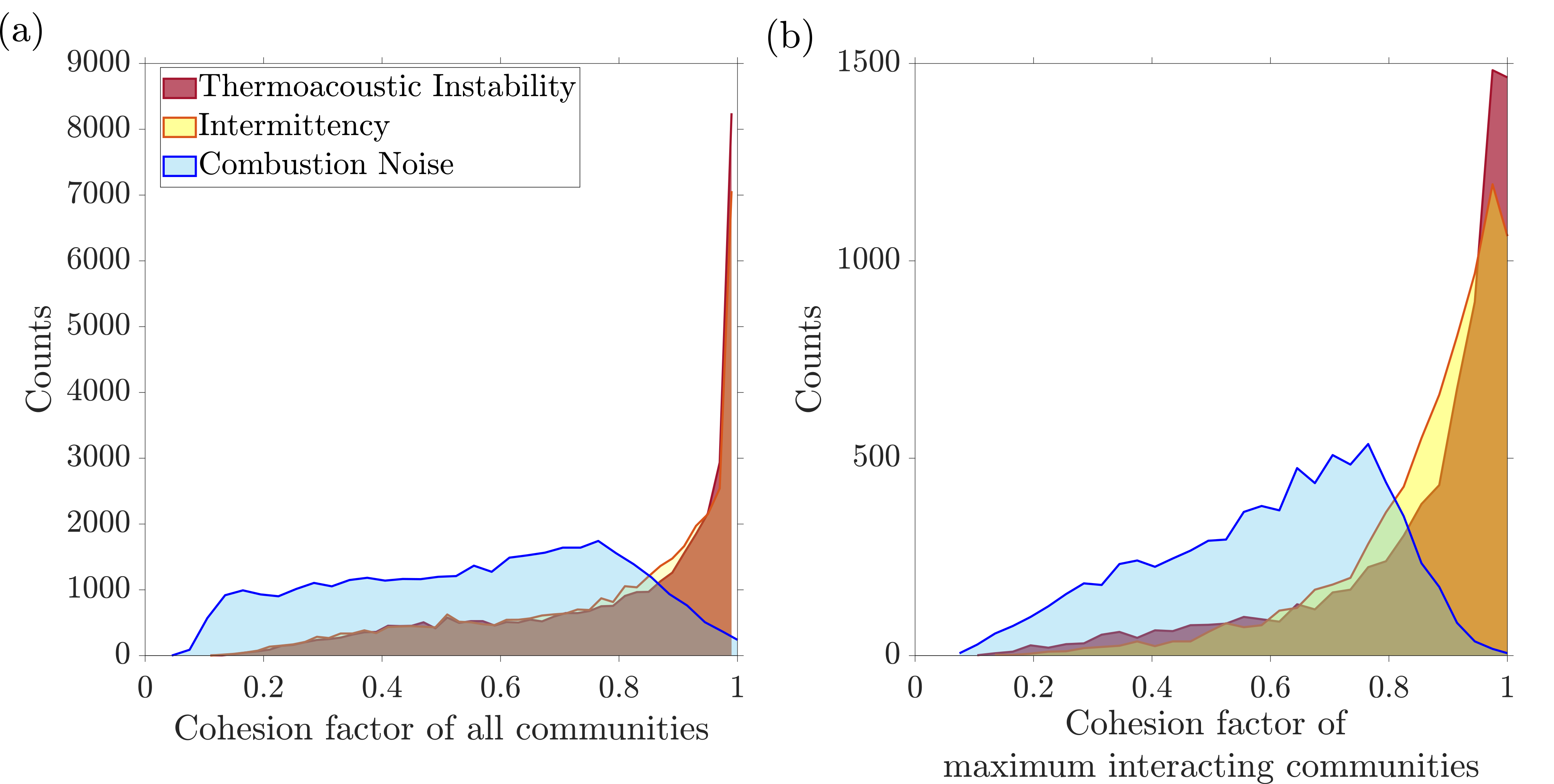}}
  \caption{Distributions of cohesion factor indicating the spatial coherence of (a) all vortical communities, and (b) vortical communities with maximum inter-community interactions during the three dynamical states.}
\label{fig_app3}
\end{figure}

To quantify the extent of fragmentation of vortical communities during the three dynamical states, we adopt a measure used by \cite{jadhav2022randomness} that measures the size of the largest 4-connected cluster in any detected community normalized by the total size of the community. This measure is called the cohesion factor. A cohesion factor equal to one implies that the communities are 4-connected everywhere, with no fragmentation at all. A cohesion factor of zero indicates an absence of any connected component. 

In figure \ref{fig_app3}, we plot the distributions of the cohesion factors of all the vortical communities and of the maximum-interacting vortical communities during the three dynamical states. We notice that the cohesion factors of the communities during the state of combustion noise are comparatively low compared with the cohesion factors observed during the other dynamical states of intermittency and thermoacoustic instability. The highly fragmented nature of vortical communities detected during the state of combustion noise can be attributed to the presence of very few small vortices scattered throughout the spatial domain. Moreover, the fragmented nature of maximum-interacting vortical communities is due to the small coherent structures primarily shed from the tip of the bluff body during the state of combustion noise. 

During the states of intermittency and thermoacoustic instability, most of the coherent structures are shed from the backward-facing step. Since these coherent structures are highly organized in the spatial domain, the vortical communities detected during these states are comparatively less fragmented than those present during the state of combustion noise. Thus, the cohesion factors of the vortical communities present during the states of intermittency and thermoacoustic instability are quite high and very close to 1.

\section{Sensitivity of community detection and associated results to \(\alpha\) values}\label{AppendixD}

We set $\alpha = 0$ to account for directed interactions between vortical elements. Choosing $\alpha = 1$ would yield similar results, as defining $u_{i \rightarrow j}$ as $A_{ij}$ or $A_{ji}$ is a matter of convention. In our formulation, we use $\alpha = 0$ to follow the convention of denoting $u_{i \rightarrow j}$ as $A_{ji}$, which is consistent with studies in physical sciences. The same notation is followed in \cite{newman2018networks}.

The sensitivity of the results to unidirectional coupling (parameterized by $\alpha$) can be explained by studying the impact of $\alpha$ on the following aspects of vortical networks:
\subsection{Community affiliations}
To study the sensitivity of results to changes in the values of $\alpha$, we calculate inter-community interactions, community affiliations, and sizes of the communities with maximum inter-community interactions for five different values of $\alpha$ - $0$, $0.25$, $0.50$, $0.75$, and $1.00$.

\begin{figure}
    \captionsetup{width=1\textwidth}
    \centerline{\includegraphics[width = 1\linewidth]{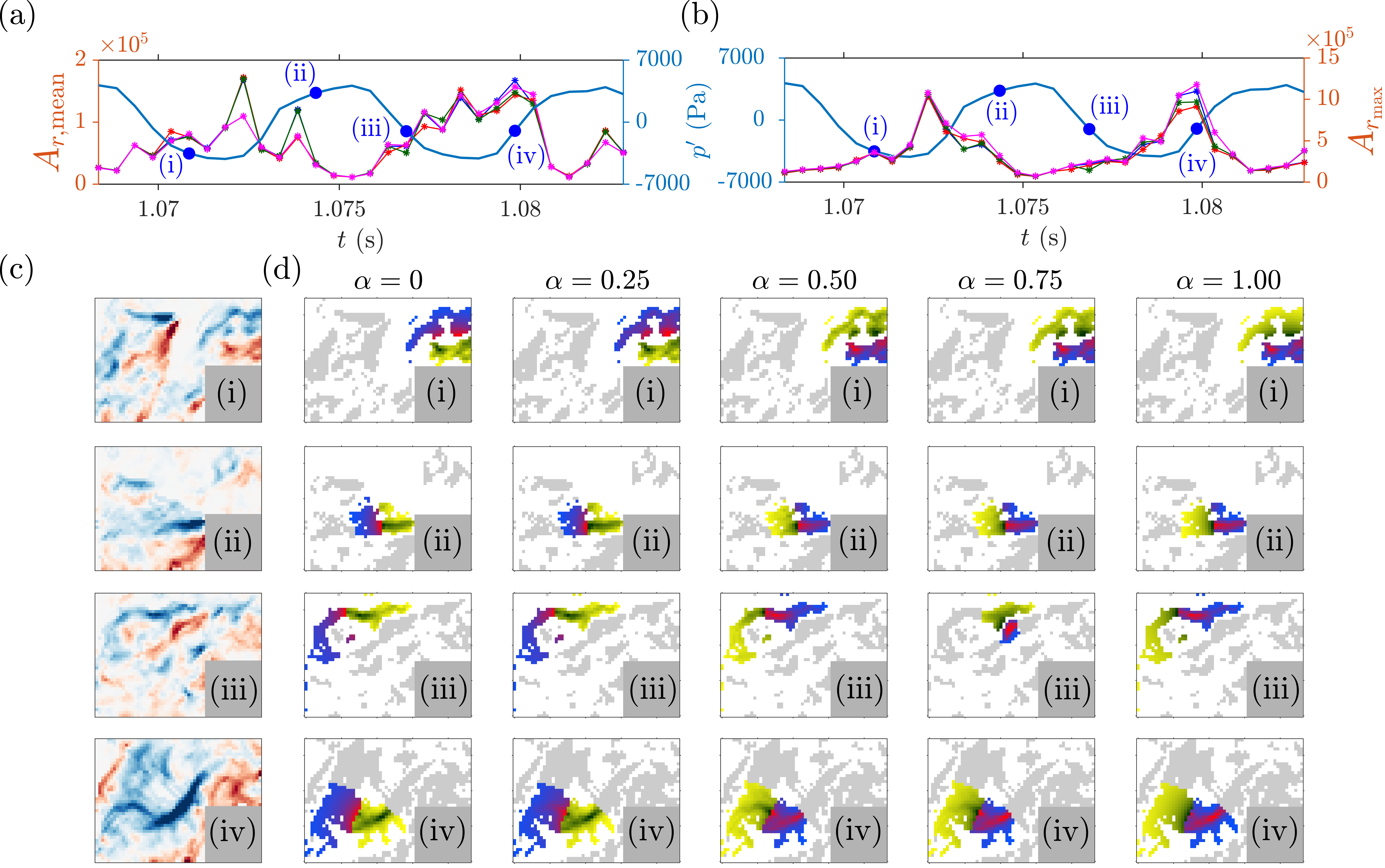}}
    \caption{Time series showing the variation of the (a) mean ($A_{r,\text{mean}}$) and (b) maximum ($A_{r,\text{max}}$) of all inter-community interactions during the state of thermoacoustic instability. The acoustic pressure fluctuations are shown in blue, and the inter-community interaction measures are marked with star-shaped markers. Five different values of $\alpha = 0$, $0.25$, $0.5$, $0.75$, and $1.00$ are used to plot $A_{r,\text{mean}}$ and $A_{r,\text{max}}$ in (a) and (b). The results indicate that the inter-community interaction measures $A_{r,\text{mean}}$ and $A_{r,\text{max}}$ do not change significantly with variations in $\alpha$. (c) Shows the vorticity fields corresponding to the four data points marked i–iv in (a) and (b). For these data points, each column in (d) illustrates the locations and strength distributions of the communities with the largest inter-community interaction for different values of $\alpha$. The pair of communities with the largest inter-community interaction remains consistent as $\alpha$ changes. However, the direction of links between the two communities reverses as $\alpha$ transitions from $0$ to $1$. This reversal is evident from the opposite colour schemes of the communities with in-degrees and out-degrees for $\alpha = 0$ and $1$.}
\label{fig_app4_1}
\end{figure}

In figure \ref{fig_app4_1}(a-b), the time series with star markers show the variation of the mean ($A_{r,\text{mean}}$) and maximum ($A_{r,\text{max}}$) of inter-community interactions for the different values of $\alpha$ during the state of thermoacoustic instability. Changing the $\alpha$ values modifies the induced velocities between vortical elements. However, this modification occurs uniformly across all pairs of nodes. Thus, the location of local maxima of network measures does not change significantly as seen in figure \ref{fig_app4_1} (a-b), although we do observe slight changes in the magnitudes of $A_{r,\text{mean}}$ and $A_{r,\text{max}}$.

To study the effect of $\alpha$ on the spatio-temporal dynamics of the maximum-interacting pair of communities, we plot the locations of the pair of communities with maximum inter-community interactions for four different data points marked (i) - (iv) in figure \ref{fig_app4_1} (a-b). The vorticity field for the four data points marked is shown in (c), and each column of (d) illustrates the maximum-interacting pair of communities for different $\alpha$ values. The two different colour schemes denote the in- and out-strength distribution of the inter-community edge weights.

The pair of communities with the largest inter-community interaction remains consistent as $\alpha$ changes from $0$ to $1$. However, the direction of links between the two communities reverses as $\alpha$ changes. This occurs since $u_{i \rightarrow j} = A_{ji}$ for $\alpha = 0$, and $u_{i \rightarrow j} = A_{ij}$ for $\alpha = 1$. This reversal is evident from the opposite colour schemes of the two interacting communities for $\alpha = 0$ and $\alpha = 1$. \\

\subsection{Phase lag between $p^\prime$ oscillations and network measures}
The phase lag shows minor sensitivity to variations in $\alpha$ since the statistics of inter-community interactions, such as their mean ($A_{r,\text{mean}}$) and maximum ($A_{r,\text{max}}$) values, change only in magnitude. The temporal patterns of inter-community interactions, that is, the time instants at which these interactions reach their local minima or maxima, remain consistent as the effect of changing $\alpha$ on the network weights occurs uniformly through the vortical network at each time instant.

\begin{figure}
    \captionsetup{width=1\textwidth}
    \centerline{\includegraphics[width = 1\linewidth]{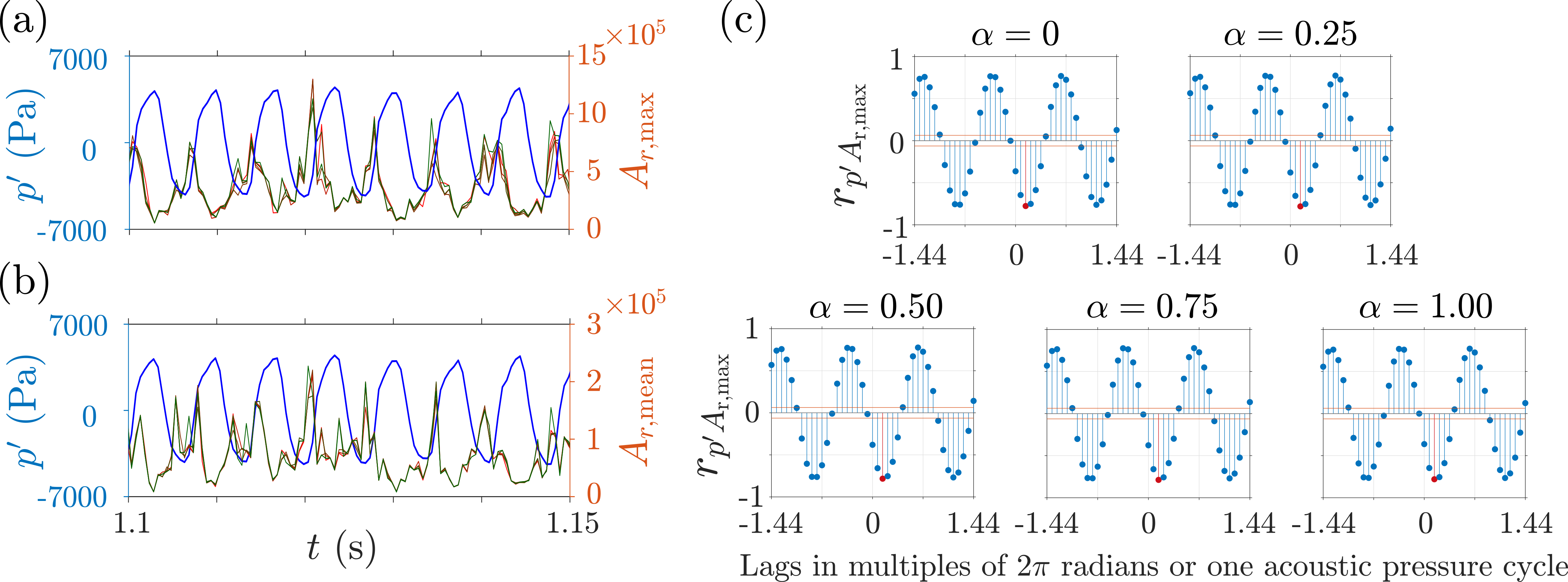}}
    \caption{Time series showing the variation of the (a) mean ($A_{r,\text{mean}}$) and (b) maximum ($A_{r,\text{max}}$) of all inter-community interactions during the state of thermoacoustic instability. The acoustic pressure fluctuations are shown in blue. (c) Shows the cross-correlation between $p^\prime$ and $A_\text{r,max}$, denoted by $r_{p^\prime A_\text{r,max}}$ and the corresponding lags in terms of multiples of acoustic pressure cycles ($2\pi$ rad is equivalent to one acoustic pressure cycle) and degrees. The minima of $r_{p^\prime A_\text{r,max}}$ is at $52 \degree$, indicating the lag between the occurrences of $p^\prime$ minima (i.e., vortex shedding from the backward-facing step) and maximum of vortical interactions. We observe that the phase lag of $52 \degree$ is independent of the value of $\alpha$ used to construct vortical networks.}
    \label{fig_app4_2}
\end{figure}

\begin{figure}
    \captionsetup{width=0.9\textwidth}
        \centerline{\includegraphics[width = 0.85\linewidth]{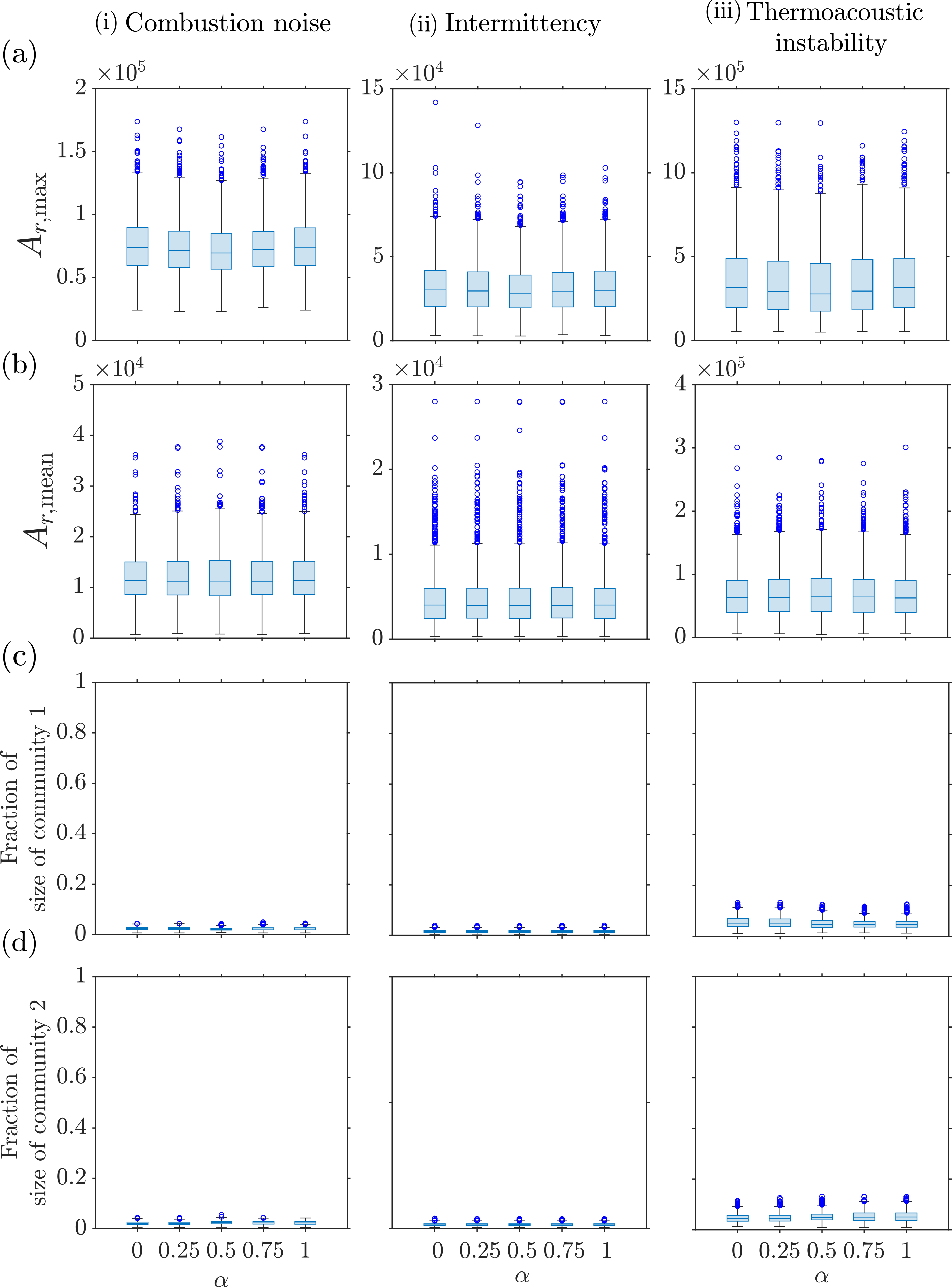}}
        \caption{Panels (a) and (b) show the box plots of $A_\text{r,max}$ and $A_\text{r,mean}$ for 1000 data points calculated for five $\alpha$ values during the states of (i) combustion noise, (ii) intermittency, and (iii) thermoacoustic instability. We observe minute changes in the distribution of the inter-community interaction measures as $\alpha$ is varied. Panels (c) and (d) show the variation of the fraction of the size of the two communities with maximum inter-community interaction with a change in the value of $\alpha$. The consistent low value of this fraction indicates the low sensitivity of the sizes of communities with maximum inter-community interaction towards different $\alpha$ values between $0$ and $1$.}
    \label{fig_app4_3}
\end{figure}

Figure \ref{fig_app4_2}(a-b) shows the variation of acoustic pressure oscillations and the mean and the maximum of all inter-community interactions for 100 time instants. The phase lag remains almost constant for all $\alpha$ values. This is further corroborated in figure \ref{fig_app4_2}(c) where we plot the cross-correlations between $p^\prime$ and $A_{r,\text{max}}$ for different values of $\alpha$. The minima of $p^\prime$ fluctuations correspond to the instants of vortex shedding. The minima of cross-correlation closest to $0 \degree$ in figure \ref{fig_app4_2}(c) indicates the phase lag where $p^\prime$ and $A_\text{r,max}$ are maximum negatively correlated. We consider maximum negative correlation since the minima of $p^\prime$ fluctuations and maxima of $A_\text{r,max}$ are used for computing the cross-correlation. We observe a consistent phase lag of about $52 \degree$ between the instants of vortex shedding and the maximum inter-community interactions for all the values of $\alpha$ considered.

\subsection{Statistics of inter-community interactions and sizes of communities with the largest interaction}
Finally, we look at the effect of $\alpha$ on the distribution of the mean and maximum of inter-community interactions and on the size of the two communities with the maximum inter-community interaction.

Figure \ref{fig_app4_3}(a,b) shows the variation of the distribution of the inter-community interaction measures calculated for 1000 data points present during all the three dynamical states with a change in the values of $\alpha$ considered while constructing vortical networks. We observe negligible changes in the distributions of $A_\text{r,max}$ and $A_\text{r,mean}$, indicating that the network measures do not vary significantly as $\alpha$ is varied from $0$ to $1$. Similarly, the size of the communities with maximum inter-community interaction (named \say{community 1} and \say{community 2} in figure \ref{fig_app4_3}(c,d)) does not change significantly with a change in the values of $\alpha$, as seen in figure \ref{fig_app4_3}(c,d).

Thus, we infer that the variation of $\alpha$ does not significantly vary the results shown in the present study. However, we emphasize using $\alpha = 0$ to maintain the unidirectional (directed) influencing capability of the vortical structures.

\bibliographystyle{apsrev4-2-titles}
\bibliography{aipsamp.bib}

\end{document}